# Trade, Trees, and Lives[*]

Xinming Du

Lei Li

Eric Zou

November 2024

### Abstract

This paper shows a cascading mechanism through which international trade-induced deforestation results in a decline of health outcomes in cities distant from where trade activities occur. We examine Brazil, which has ramped up agricultural export over the last two decades to meet rising global demand. Using a shift-share research design, we first show that export shocks cause substantial local agricultural expansion and a virtual one-for-one decline in forest cover. We then construct a dynamic area-of-effect model that predicts where atmospheric changes should be felt – due to loss of forests that would otherwise serve to filter out and absorb air pollutants as they travel – downwind of the deforestation areas. Leveraging quasi-random variation in these atmospheric connections, we establish a causal link between deforestation upstream and subsequent rises in air pollution and premature deaths downstream, with the mortality effects predominantly driven by cardiovascular and respiratory causes. Our estimates reveal a large telecoupled health externality of trade deforestation: over 700,000 premature deaths in Brazil over the past two decades. This equates to $0.18 loss in statistical life value per $1 agricultural exports over the study period.

[*] Du: Department of Economics, National University of Singapore (email: xdu@nus.edu.sg); Li: Faculty of Business and Economics, University of Göttingen (email: lei.li@uni-goettingen.de); Zou: Ross School of Business, University of Michigan and NBER (email: ericzou@umich.edu). We thank Philipp Ager, David Autor, Ricardo Dahis, David Dorn, Rafael Dix-Carneido, Thiemo Fetzer, Lisandra Flach, Bård Harstad, Karen Helene Ulltveit-Moeandre, Andreas Moxnes, Jan Schymik, Arthur Seibold, Joseph Shapiro, Camille Urvoy, Andreas Kotsadam, Ulrich Wagner, Kathrine von Graevenitz, and participants at various seminars for helpful comments. Zihao Chen, Guilherme Lindenmeyer, and Raphaël Pérot provided excellent research assistance. All errors are our own.

# 1. Introduction

Globalization creates complex interactions between consumption, production, and environmental impacts that can occur in locations far apart.[1] This paper demonstrates the importance of such telecoupling for understanding the health burden of international trade. We study Brazil, where agricultural export value has quadrupled over the last two decades due to rising global demand. This surge is accompanied by substantial forest losses to accommodate the agricultural expansion. We show that these export-driven deforestations have far-reaching health ramifications, as the lost forests would otherwise serve to absorb harmful air pollutants as they travel, influencing air quality in cities hundreds or even thousands of miles downstream. We find that the magnitude of the health effects is substantial, even when compared to the entire value of Brazil's agricultural trade over the study period.

We are motivated by an emerging debate on the "non-economic" effects of trade, particularly with regard to its interaction with the environment (Copeland, Shapiro, and Taylor, 2022). This is especially relevant in developing country context, where the reliance on exporting resource-intensive products plays a key role in achieving development goals (Frank and Schlenker, 2016). In current literature, the environmental costs of deforestation are predominantly associated with climate change and disruptions to ecosystems (Copeland, Shapiro, and Taylor, 2022; Balboni et al., 2023). These include projected alterations in global temperature and precipitation patterns (Bonan, 2008), impacts on biodiversity (Dasgupta, 2021), and increased risk of crossing ecosystems tipping point (Franklin and Pindyck, 2018). Though expected to be profoundly important, these effects are hard to empirically pin down due to their long-run nature. Our paper instead examines human health effects that can be directly measured in the data *today*. The main message of our work is that the environmental costs of trade-induced deforestation are considerable, even when considering immediate cardio-respiratory health impacts alone.

---

[1] Economists have extensively studied how trade enhances growth (Grossman and Helpman, 1990) and productivity (Alcalá and Ciccone, 2004), and how it interacts with institutional quality (Levchenko, 2007). Empirical studies show international trade could affect the environment in positive or negative ways (e.g., Antweiler et al., 2001; Frankel and Rose, 2005; Managi et al., 2009; Abman and Lundberg, 2020), and emphasize the large spatial ranges of these impacts. Incorporating multiple locations and environmental consequences is important in assessing the overall welfare effects of globalization.

We begin by estimating the causal effect of agricultural exports on local forest land coverage. During the two decades of our study period (1997–2019), Brazil significantly expanded its agricultural land use, coinciding with a substantial decline in forest cover. This pattern of deforestation is widely suspected to be associated with growing agricultural export demands, particularly from China and the European Union, which drive farmers to employ slash-and-burn methods for rapid land clearing to boost agricultural production. We estimate the causal link between agricultural exports and deforestation using a shift-share research design, which exploits variation in regions' exposure to global demand shocks ("shift") due to historical differences in export capacity across product categories ("share"). Brazilian regions are exposed to idiosyncratic import shocks across a wide range of product categories, providing an apt context for applying the shift-share research design (e.g., Adao, Kolesár, and Morales, 2019; Borusyak, Hull, and Jaravel, 2022). Our estimates show that each 1,000 BRL increase in export per capita reduces forest cover in the area by 0.174 percentage points. Using the same econometric framework, we demonstrate that areas experiencing forest retreat also saw agricultural expansion. Consistent with existing knowledge, this effect is mainly explained by increases in soybean and sugar farming—both land-intensive crops long associated with deforestation concerns.

Next, we estimate the impact of forest loss on downstream environmental and health outcomes. To accomplish this, we need to identify the extent of cross-boundary pollution spillover: the impact of a pollution source—or in our case, the absence of forests that would have reduced pollution—does not conform to administrative boundaries and can extend far beyond its origin (e.g., Heo, Ito, and Kotamarthi, 2023). We develop a simple area-of-effect (AoE) model, where we predict the geo-temporal influence of a given city's forest change by solving a dynamic model that tracks airflows across space and time. From this model, we obtain information on the intensities of upwind-downwind linkages, or the lack thereof, of all pairs of cities in Brazil in all months. Our model is simplistic enough to be computationally feasible, but at the same time it is sufficiently realistic: we show that the model's output linearly predicts the correlation between upwind and downwind cities' observed pollution levels (which we did *not* use in training the model); conversely, when our model predicts a lack of up-downwind link, there is indeed a lack of correlation between two cities' pollution levels. Put differently, our model appears to predict the potential extent of pollution transport very well.[2]

Building on this data, we develop a quasi-experimental design to estimate the impact of forest losses on downwind environmental quality and health. For each "sender" city (where deforestation happens), potential "receiver" city (those on the downwind path of the sender city), and month, we examine the relationship between the sender's forest volume (acreage) and the receiver's environmental and health outcomes, allowing this relationship to vary according to the intensity of the sender-receiver downwind intensity score as suggested by the AoE model above. Our estimation includes flexible controls such as sender-receiver city pair by month-of-year fixed effects and year fixed effects, which enables us to leverage quasi-random wind patterns to isolate how upwind forests influence downwind conditions.

Our findings show that deforestation in upwind cities significantly leads to increased air pollution and higher mortality rates in downwind cities. The mortality effect is primarily driven by excess deaths from cardiovascular and respiratory causes, consistent with a pollution exposure pathway. In placebo tests, we find that deforestation has no effect on plausibly irrelevant mortality outcomes, such as accident-related deaths. We also examine periods of calm wind conditions and show that the pollution and mortality effects do not arise when the same city pairs lack a significant upwind/downwind relationship.

Losses in upwind forests can lead to changes in downwind air quality and health through two pathways. The first is a pollution cleansing effect, where deforestation reduces the forest's capacity to filter and absorb airborne pollutants, which would otherwise help cleanse the air before it reaches downwind areas. The second is a fire effect, where deforestation, often done using slash-and-burn method, increases the frequency of fires, producing smoke that drifts toward downwind cities. To disentangle these effects, we use remote-sensing data on fire incidents and augment our estimation model to allow fires to have flexible and independent effects on downwind conditions. Our analysis shows that the pollution and mortality estimates remain largely unchanged with the inclusion of fire information, indicating that the fire pathway plays a minor role.

---

[2] An expanding body of literature has leveraged variability in wind conditions to estimate the causal effects of air pollution (e.g., Deryugina et al., 2019; Rangel and Vogl, 2019; Anderson, 2020; Graff Zivin et al., 2023; Guidetti, Pereda, and Severnini, 2024). A key methodological innovation in our work is the development of a computationally feasible approach to build a comprehensive matrix of wind transport relationships across city pairs. This is crucial when the researcher's goal is not only to measure the burden of pollution in the destination area but also to trace that burden back to its sources.

Together, these estimates can be readily integrated to compute the overall impact of trade-induced forest losses. Our calculation reveals that agricultural exports over the study period are responsible for over 3.6 million hectares of forest losses and 732,000 premature deaths. This is a substantial figure compared to the total number of deaths reported in the entire country in 2018, which is 1.28 million. Using a Value of Statistical Life (VSL) approach, we calculate that the life value loss associated with these excess deaths is around 513 billion USD. This is about 18 percent of the total agricultural value of Brazil over the entire study period – thus our headline number. This estimate is likely conservative, as our study focuses on monthly mortality responses, and the literature indicates that the long-term effects of pollution are usually more significant (Ebenstein et al., 2017). Moreover, individuals who do not die from pollution exposure may still experience other significant, albeit non-fatal, consequences, such as reduced productivity (Graff Zivin and Neidell, 2012; Borgschulte, Molitor, and Zou, 2022).

We make three contributions to the literature. First, we quantify the causal impact of trade on deforestation. While the trade and environment literature has traditionally focused on the role of industrial emissions, there is a growing body of work that examines the environmental impact of trade through its interaction with natural resources. To the best of our knowledge, this paper provides among the first causal estimates on the impact of exports on deforestation. The most closely related work is Carreira, Costa, and Pessoa (2024), who study the impact of agricultural productivity changes and China shocks – and their interactions – on forest cover in Brazil.[3] Earlier work such as Ferreira (2004) also documents a negative relationship between trade and forest cover, though mostly based on cross-country comparison without causal identification.[4]

---

[3] Carreira, Costa, and Pessoa (2024) identify a strong negative impact of new agrotechnology (genetically engineered soy seeds) on forest cover in Brazil, along with a negative, though less statistically robust, effect from trade shocks. Our findings on trade-driven deforestation in Section 4 align broadly with their results, though our approach differs in two key ways. First, instead of focusing solely on demand shocks from China, our instrumental variable captures agricultural demand variation from multiple major importing countries. Second, we utilize a more extensive dataset with additional years and refined satellite measurements of forest cover, likely enhancing the precision of our estimates. In unreported analysis, we successfully replicate Carreira, Costa, and Pessoa (2024) using same subsample, satellite measurement, and IV construction.

[4] The general link between agricultural expansion and deforestation is more extensively studied. See, for example, Busch and Ferretti-Gallon (2017), Heilmayr et al. (2020), and Penderill et al. (2022).

Second, we add to the literature on the environmental and health effects of forests. This literature is nascent in itself, but emerging applications have found important implications of forest presence on weather, climate, agriculture, infectious diseases, health, among other social economic outcomes such as property values (Berazneva and Byker, 2017, 2022; Garg, 2019; Jones and Goodkind, 2019; Druckenmiller, 2020; Han et al., 2021; Arujo, 2023; Grosset, Papp, and Taylor, 2023; Li, 2023; Xing et al., 2023). The most closely related study is Xing et al. (2024), which studies the impact of an urban afforestation program in Beijing, China, on air quality and respiratory emergencies in the city. Our study is the first to identify a direct link between forests and mortality rates, adding a significant dimension to the costs associated with forest loss, and doing so across a wide, inter-city geographical scale.

Third, and most importantly, our study connects the dots between trade, deforestation, and health, revealing natural resource depletion as a critical yet overlooked pathway through which trade impacts health. Trade shocks are understood to affect health via either an income channel or an emission channel. Trade-induced job loss has been shown to increase drug overdoses (Pierce and Schott, 2020), reduce access to employment insurance (Guerrico, 2021), and worsen workers' physical and mental health (Adda and Fawaz, 2020). Bombardini and Li (2020) show that Chinese cities specializing in dirty industries experience both pollution and income effects, leading to increases in mortality. In contrast, cities specializing in clean industries only enjoy income effects and experience decreases in mortality. Gong et al. (2023) examine how foreign demand shocks due to the global economic crisis reduce demand for Chinese products, drive local changes in air quality, and affect mortalities. In contrast, our study emphasizes the role of natural resource—in our case, the Amazon forests—which we demonstrate has a significant contribution to human health production. Trade-induced depletion of this natural resource can substantially undermine the social value of trade.

More broadly, our work provides a quantitative basis to discuss the environmental damage of agricultural trade as well as the value of forests as an important piece of natural capital. Domestic environmental regulations are among the limited number of tools the government can use to mitigate the impacts of trade on natural capital depletion (Copeland, Shapiro, and Taylor, 2022). Such policymaking must be grounded in rigorous economic science that monetizes the value of natural assets, and our study describes a feasible framework to achieve this.[5] In addition to providing an overall estimate of the export-health trade-off, our estimates may also be useful for targeting, e.g., identifying areas where the same amount of deforestation would lead to the largest downwind environmental health damage.

Section 2 continues with institutional knowledge on trade, deforestation, the environment, and human health. Section 3 describes the overall research framework and data sources. Section 4 presents the causal estimation of the effect of export on forest coverage. Section 5 presents the health analysis. Section 6 presents the cost calculation and concludes the paper.

## 2. Institutional Background

### 2.1 Agricultural Export and Deforestation in Brazil

Brazil is the eighth largest economy and the world's seventh most populous country. At the same time, Brazil is an ecological hotspot, home to 60 percent of the Amazon rainforest and the Cerrado biome, the world's most biodiverse tropical savannah. It is estimated that Brazil holds about 10 percent of the world's known biodiversity (Lewinsohn and Prado, 2005). In the shadow of the rapid expansion of agribusiness in Brazil, deforestation is emerging as an environmental crisis. In many developing countries, including Brazil, farmers are driven to deforest land as a means to increase their agricultural output and income. The prevalent method of land conversion involves slash-and-burn techniques, where trees are cut down and the remaining vegetation is burned, enriching the soil with nutrients from the ash. The opening of new lands for agriculture is frequently facilitated by government subsidies or lax enforcement of environmental regulations, further incentivizing farmers to engage in these practices (Fearnside, 2005).

The interplay between agricultural expansion and deforestation is particularly evident in the Brazilian Amazon Rainforest and the Cerrado, where the cultivation of soybeans and the expansion of cattle ranching are the primary agricultural activities driving deforestation (Song et al., 2021). Soybeans, a major export crop, have seen increased demand globally, particularly from China and the European Union, for use as livestock feed and in various food products. This demand has propelled Brazilian farmers to clear vast tracts of the Amazon to cultivate this lucrative crop. Similarly, cattle ranching, which requires extensive land area for grazing, has expanded dramatically, making Brazil one of the world's largest beef exporters. The profitability of these agricultural ventures, along with government policies that frequently favor economic growth over environmental protection, is widely regarded as a major driver of significant deforestation in the region (Barona et al., 2010).

---

[5] See, for example, Recommendation 3 of *National Strategy to Develop Statistics for Environmental-Economic Decisions*, Office of Science and Technology Policy, Office of Management and Budget, Department of Commerce (2023). In addition to direct regulations, many alternative methods for fighting deforestation may also benefit from concrete estimates of health effects of forest losses, such as conservation contracting and ecosystem payment programs (Jayachandran et al., 2017; Harstad, 2024).

## 2.2 Forests and the Environment

Forests represent a crucial component of natural capital, contributing to the ecosystem and the environment. They function as indispensable habitats for 80% of the global terrestrial species (Aerts and Honnay, 2011). Besides, forests contribute to water, oxygen, and carbon cycles, and other biogeochemical balances. They also play a pivotal role in the global and local climates. Forests are the world's largest storehouses of carbon. Trees absorb greenhouse gasses, and are primary instruments for carbon storage featured in carbon offset programs like the REDD+. At the local level, forests affect micro-climate via shading and evaporation (Bonan, 2008; Pan et al., 2011).

Forests matter a lot for human well-being. Over 1.6 billion people depend on forests for food, fuel, or shelter, and 70 million people including indigenous communities rely on forests for settlement (UN, 2021). Forests also mitigate damages from pollution and natural disasters such as floods. In this paper, we focus on air pollution impacts of forests as short-term consequences of trade-induced deforestation. Unlike other effects, air pollution responds quickly to tree cover change and could be linked to export changes of interest. Air pollution also leads to immediate mortality responses, enabling us to quantify health externality.

Why do forests mitigate air pollution? First, leaves and bark serve as filters for particles, providing deposition sites for airborne pollutants. Previous studies have documented that forests are efficient pollution collectors and improve air quality (e.g., Betts et al., 2008; Janhall, 2015). Second, leaves can absorb specific hazardous gases like $SO_2$, NOx, and $O_3$. They can also interact with pollutants and alter atmospheric compositions. Leaves are reported to biodegrade or transform pollutants into less or nontoxic molecules (Wei et al., 2017). Third, trees indirectly affect the formation and deposition of air pollution by altering the micro-climate. Ozone formation is inhibited by lower temperatures (Coates et al., 2016), and increased rainfall decreases the lifespan of particles.

The link between deforestation, air pollution, and health can span a wide spatial range due to the long-distance transport of pollutants through wind flows. Winds carry upwind pollution into downwind regions, thus wind speed, direction, and other climatic factors determine the residence time and location of pollution. Existing papers have shown the impacts of cross-regional or transboundary air pollution on downwind areas (e.g., Heo, Ito, and Kotamarthi, 2023). Focusing on trees, policymakers have implemented afforestation initiatives to protect downwind populations from air pollution. For example, in China, the Three-North Shelter Forest is designed to restrain sandstorms from the Gobi Desert and improve air quality near the capital region.

## 3. Research Framework and Data

### 3.1 Research Framework

Before proceeding, it is worth laying out how various parts of our empirical estimation are eventually going to line up. Our goal is to estimate the external health effects of trade-induced deforestation that occurs in city i, recognizing that this effect need not confine locally to where deforestation occurs. The empirical analyses can be thought of as being organized around the following conceptual equation:

$$\text{Health Effects}_i = \sum_r \underbrace{\frac{\partial \text{Forest}_i}{\partial \text{Trade}_i}}_{\text{Section 4}} \cdot \Big( \underbrace{\frac{\partial \text{Health}_r}{\partial \text{Forest}_i}}_{\text{Section 5.2}} \mid \overbrace{\text{Wind}_{i \to r}}^{\text{Section 5.1}} \Big) \quad (1)$$

Section 4 estimates the causal effect of trade on deforestation ($\frac{\partial \text{Forest}_i}{\partial \text{Trade}_i}$). This estimate is entirely local in nature: we want to learn how much of city i's deforestation is due to trade shocks to *that* city. Section 5 estimates the spillover health effects of city i's deforestation on all potential "receiver" cities indexed by r. Section 5.1 first builds an algorithm that identifies these potential receivers. We use high-frequency wind direction information to create a downwind exposure index $\text{Wind}_{i \to r}$ which summarizes the degree to which city $i$ may have an influence on city $r$'s environmental condition due to atmospheric motions. Section 5.2 then builds on this information to estimate the relationship between city $i$'s forest cover and receiver city $r$'s health outcomes – which we show is highly dependent on, and nonlinear with respect to, the downwind exposure index. The total effect of city $i$'s is thus the summation across its effects on various downwind receiver cities.

In Section 6, we come back to an empirical version of equation (1) to derive our headline, cost-of-trade-induced-deforestation number.

### 3.2 Data

Our analysis is done almost entirely using publicly available datasets, many of which are maintained by the Brazilian government. We briefly describe each source.

**Exports.** Our export data is from the Comex database provided by Brazil's Foreign Trade Secretariat (Secretaria de Comércio Exterior - Comex). The original data is at the municipality by month level and is structured using the Harmonized System at the 4-digit level ("HS-4"), providing detailed categorization of exported goods. We have access to data between 1997 and 2019.

**Land Use.** Our land use data is from MapBiomas. Based on the Landsat satellite images produced by NASA and USGS, MapBiomas processes the pixel-per-pixel data with machine learning algorithms through the Google Earth Engine platform. It produces annual land use and land cover maps and provides information on the dynamics of different types of land cover for Brazil from 1985 to 2022 on a 30m spatial resolution. Specifically, there are six types of land, namely forest, non-forest natural formation, farming, non-vegetated area, water, and not observed. We mainly analyze the data on forest land and farming land. There are four types of forest lands, namely forest formation, savanna formation, mangrove, and sandy coastal plain vegetation. There are four types of farming land, namely pasture, agriculture, forest plantation, and mosaic of uses. For agricultural land, the MapBiomas offers information on major crops, such as soybean, sugar cane, rice, cotton, coffee, and citrus. The data used in our paper is the 7.0 version obtained in 2022.

**Meteorology.** We draw daily temperature and precipitation data from BDMEP (Banco de Dados meteorológicos para Ensino e Pesquisa). During the majority of our study period (2000-2015), BDMEP provided data from 150 weather monitoring stations. The network expanded to include 5,611 stations for the years 2016-2020. For each station-day, we observe daily minimum and maximum temperatures (°C) – from which we calculate average temperature by taking the average of the two – and precipitation (mm). We then compute station by month average temperature and precipitation. To derive city-level weather variables, we match all weather stations within a 300-kilometer radius of each city's centroid and calculate the average temperature and precipitation across these matched stations.

We use wind direction and speed data from the European Centre for Medium-Range Weather Forecasts Reanalysis 5 (ERA5) reanalysis product. This dataset provides hourly U (horizontal) and V (vertical) components of the wind vectors at a 0.25° grid resolution. We use this data to compute daily prevailing wind directions, which are used in the air flow analysis in Section 5.

**Air Quality.** Air quality data is sourced from IEMA (the Instituto de Energia e Meio Ambiente). It contains hourly air quality monitoring data from 380 stations in 28 cities from 2015-2022. The IEMA data keeps track of six different air pollutants: fine particulate matter ($PM_{2.5}$), coarse particulate matter ($PM_{10}$), ozone ($O_3$), nitrogen dioxide ($NO_2$), sulfur dioxide ($SO_2$) and carbon monoxide (CO). Because trees can mitigate both particulate and gaseous pollutants (Section 2.2), rather than focusing on these pollutants individually, we compute the z-scores for each of the six pollutants and then average these z-scores to construct an overall air pollution index. This index serves as our primary measure of air quality in the regression analysis. Regression results using individual pollutant concentrations as the outcome variables are available in the Online Appendix.

**Mortality.** We obtain mortality microdata from the Mortality Information System (Sistema de Informações sobre Mortalidade, or SIM) via the Brazilian National Health System Information Technology Department (Departamento de Informática do Sistema Único de Saúde, DATASUS). The SIM data are derived from administrative death records, which are completed and issued by physicians, and collected from hospitals and medical examiner offices.

From the death record microdata, we observe the decedent's municipality of residence, date of death, and the underlying cause of death coded in ICD-10 classification. Using these information, we aggregate the microdata to obtain death counts at the microregion-by-month level between 2000-2021.[6] We categorize deaths into those due to cardiovascular or respiratory causes ("cardiovascular", ICD10: I00-I99, J00-J99) and all other, non-cardiovascular causes. Separately, we count deaths due to accidents ("external", ICD10: S00-T98, V01/V989), which we will use as a "placebo" outcome in our econometric analysis. Finally, we convert death counts to death rate by dividing microregion level population counts from the Brazilian Institute of Geography and Statistics.

**Geography.** Because municipality boundaries changed over time over our study period, we aggregate geographic units to the 3,826 Minimum Comparable Areas (henceforth "AMC" area) level for temporal consistency (Ehrl, 2017). We drop five municipalities that cannot be identified in the municipality-to-AMC crosswalk.[7] Our primary trade-deforestation analysis is based on AMC-level data. Our environmental data have a coarser resolution than AMCs (e.g., 0.25 degree for wind, 380 air quality stations). Therefore, in airflow modeling and the subsequent environmental health analysis, we aggregate AMCs into 557 "microregions" as defined by the Brazilian Institute of Geography and Statistics (IBGE).

## 4. Trade, Agricultural Expansion, and Deforestation

### 4.1 Raw Statistics

We first present a set of summary statistics on the shifts in Brazil's trade structure, agricultural expansion, and deforestation. Figure 1, panel (a) plots Brazil's national total agricultural export value, which grew from about 20 billion USD in 1997 to over 80 billion USD in 2018. Appendix Figure 1 panel (a) shows distribution of export values across agricultural products. Appendix Figure 1 panel (b) shows that the increase in exports is mainly driven by rising demand from China and the EU.[8] Figure 1, panel (b) shows the decline in Brazil's forest cover over the past decades, accompanied by a virtual one-for-one increases in farmland. The map on the right side of the panel highlights areas of deforestation which shows a strong spatial correspondence with regions experiencing significant agricultural export growth.[9] We next describe our econometric method to quantify the causal effect of export shocks on deforestation.

---

[6] The SIM mortality data began in 1997, although we drop data from 1997 to 1999 due to missing information on causes of death.
[7] Mojuí dos Campos, Pescaria Brava, Balneário Rincão, Pinito Bandeira, and Paraíso das Águas.

## 4.2 Research Design

The main estimation equation is:

$$\Delta \text{Land Cover}_{i,y} = \beta \cdot \Delta \text{Export}_{i,y} + \lambda \cdot W_{i,y} + \varepsilon_{i,y} \quad (2)$$

where $\Delta \text{Land Cover}_{i,y}$ is the four-year rate of change in a certain type of land cover at AMC area $i$ and year $y$. For example, when the outcome of interest is deforestation, the land cover variable is defined as $\Delta \text{Forest}_{i,y} = (\text{Forest}_{i,y+4} - \text{Forest}_{i,y})/\text{Land}_{i,y}$ where "Forest" denotes forest cover in acreage in the area-year, and "Land" is total area land size. We use a four-year differencing period in both equation (2) and in the construction of our instrumental variable below. In Appendix Table 3, we present robustness checks with alternative differencing windows, ranging from one to six years. We chose the four-year period as our primary specification as the magnitude of the estimated effect stabilizes from this point onward, suggesting that a four-year window is sufficiently long to capture the temporal dynamics between export shocks and deforestation activities.

The key right hand side variable $\Delta \text{Export}_{i,y}$ is the four-year change in agricultural exports per capita. $W_{i,y}$ is a series of control variables, including macro-region fixed effects, year fixed effects, and region-time control variables including income per capita, literacy rate, rural population, urban population, population density, and the presence of genetically-modified soy.[10] The coefficient of interest β therefore captures the impact of agricultural export growth on land cover changes. We cluster standard errors at the AMC level. All regressions are weighted with each area's agricultural population in year y.

---

[8] Appendix Figure 3, panel (a) shows the geographic distribution of the agricultural export growth. Not surprisingly, growth hotspots concentrate in areas with high crop production. Panel (b) reports corroborative evidence of remarkable agricultural employment growth with a similar geographic pattern. We present maps of mining product exports – the other main sector that saw substantial growth during the study period – in Appendix Figure 4. The geographic pattern is markedly different from that of agricultural exports.

[9] We report actual levels of farmland gains and forest losses in Appendix Figure 2.

Simple OLS estimation of equation (2) may suffer from endogeneity and measurement errors. We build the following shift-share instrumental variable (IV) for the export variable:

$$IV_{i,y} = \left[\sum_j \frac{\text{Export}_{i,j,t-4}}{\text{Export}_{i,t-4}} \cdot \frac{\text{Import}_{j,t+4} - \text{Import}_{j,t}}{(\text{Import}_{j,t+4} + \text{Import}_{j,t})/2}\right] \cdot \frac{\text{Export}_{i,t-4}}{\text{Pop}_{i,t-4}} \quad (3)$$

where $\text{Export}_{i,j,t-4}/\text{Export}_{i,t-4}$ is the "share" of product j in area i's exports at period $t-4$, which captures an area's historical export of a certain type of agricultural product. $\text{Import}_{j,t}$ is major importing countries' total import demand in product j and year t. The term $\frac{\text{Import}_{j,t+4} - \text{Import}_{j,t}}{(\text{Import}_{j,t+4} + \text{Import}_{j,t})/2}$ therefore captures percent change (or "shift") in product demand over the period of interest. Together, the shift-share terms in brackets of equation (3) capture the variation in an area's exposure to global demand shocks due to historical differences in export capacity across product categories. To express the IV variable in terms of per-capita dollar value – consistent with the endogenous variable in equation (2) – we multiply by $\frac{\text{Export}_{i,t-4}}{\text{Pop}_{i,t-4}}$, which represents area i's overall historical export exposure. Because the export values of many products are zero in early years, it is not possible to calculate growth for some products. To address this issue, we construct the modified version of import growth following Davis and Haltiwanger (1992), where the denominator is the average of the start-of-period value and end-of-period value.

The shift-share IV strategy relies on trade shocks being diverse, uncorrelated across product categories, and as-good-as-random conditional on shock-level unobservables and exposure weights (Borusyak, Hull, and Jaravel, 2022). Several empirical facts and tests support these assumptions.

First, Brazil's agricultural exports span a broad range of products. Between 1997 and 2019, our study period includes over 190 product categories at the HS-4 level, with an average Herfindahl–Hirschman index (HHI) of 0.07, indicating low concentration. Second, import demand shocks vary widely across products. As shown in Appendix Figure 5, import demand growth over the study period fluctuates significantly across HS-4 product categories. Together, these two points support the assumption that the shift-share variation in our IV estimation is driven by diverse import growth shocks across a wide range of products, rather than by concentrated shocks in a few product categories.

---

[10] We follow Carreira, Costa, and Pessoa (2024) in selecting these control variables. Our estimates are insensitive to the exclusion of these covariates.

The exogeneity of shocks cannot be tested directly. Instead, we conduct a balancing test following Borusyak, Hull, and Jaravel (2022). Specifically, we regress pre-shock observable characteristics on the IV variable, controlling for the same set of fixed effects variables as in equation (2). We look at control variables that we used as covariates in equation (2), as well as climate variables such as temperature, precipitation, and humidity which might correlate with agricultural suitability. Appendix Table 3 shows that the IV variable does not significantly correlate with these predetermined characteristics.

Yet another common concern with the shift-share strategy is correlated shocks, where regions with similar export structures may experience similar shocks, potentially leading to overrejection in 2SLS estimation (Adão, Kolesár, and Morales, 2019). We follow Adão, Kolesár, and Morales (2019) and conduct a placebo test, where we simulate shift-share IV variables with shift shocks randomly drawn from a normal distribution with mean 0 and variance 5. We then estimate the reduced form effect by regressing forest growth on the placebo IV, repeating this procedure 1,000 times. To assess overrejection, we examine rejection characteristics in the distribution of placebo coefficients. Appendix Figure 6 shows that the placebo regressions render significant results approximately 10.6% of the time at the 5% significance level and 1.5% of the time at the 1% significance level, suggesting a lack of severe overrejection issues.

## 4.3 Results

Figure 2 reports the primary estimates. Each node on the tree represents a separate regression. All regressions follow the exact same IV specification as laid out in equations (2)-(3), except for the outcome variables which are denoted by the name of the node. Branches of the tree represent hierarchies of the land use variables defined in the MapBiomas dataset. For example, the "forest" land use category consists of four subcategories: "forest formation", "savanna formation", "mangrove", and "sandy". The values on the four sub-nodes represent the causal impacts of export on each subcategory, which collectively sum to the overall causal impact on the main forest node.

The headline "export-deforestation" estimate of this paper is shown on the "ΔForest" node of the figure: each 1,000 BRL increase in export per capita reduces forest cover by 0.174 percentage points. This estimate is statistically significant at the 5 percent level, and the effect size converts to about 5.7 percent of a standard deviations (S.D.) reduction in forest cover per 1 SD change in export. The sub-nodes show that this total deforestation effect is driven mainly by decrease in forest and savanna formation.

Now come to the lower parts of the graph. We show that export leads to significant increases in farmland coverage by about 0.214 percentage points. This effect is diluted by the fact that, in the land use data, "forest plantation" is counted toward farming activities, which – consistent with our deforestation results above – is *negatively* affected by exports. Consistent with institutional knowledge, the main farming effect is driven by increases in agriculture by 0.334 percentage points.

We can further look at which crop types drive the agriculture effects. The right-hand-side parts of the graph show that export's effect of agriculture is driven primarily by an increase in soybean farming. This is once again consistent with the well-known fact that soybean demand from China and the EU has substantially increased over the past decades.

## 5. Health Effects

### 5.1 Area-of-Effect Modeling

To capture downwind environmental and health effects, we build a matrix that summarizes monthly wind flow intensities between all pairs of cities in Brazil. In the interest of space, here we provide an intuitive explanation of the procedure, leaving computational details to the Online Appendix.

The input data are 0.25-degree resolution by hourly wind direction and speed information (i.e., vectors) from the ERA5 product. We first generate a spatial representation of wind *flows* from individual wind *vectors*, as illustrated in Figure 3a. Beginning from a particular day and city, we construct streamlines by sequentially following the wind's speed and direction on a daily basis.

This process maps out the evolving trajectories of the wind field, giving us daily representations of the distribution of wind flows throughout the country.

Our goal is to build an index, $Wind_{i \to r,d}$, which is a summary of downwind intensity blowing from a sender city $i$ to a receiver city $r$ on day $d$. Note that, for any pair of cities that are significantly distant from each other, it is meaningless to talk about up/downwind relationship on any given "day" because winds blowing from the sender city may take days to arrive at the receiver city. We therefore track the trajectory of winds "originating" from a sender city and their impacts of downwind cities over multiple days – or "steps" as we refer to them below – using the wind streamline data that we constructed earlier.

We define downwind intensity score using the following formula:

$$Wind_{i \to r,d,t} = \exp\{-\alpha \cdot rad_t - \beta \cdot |\theta|_{i \to r,d,t} - \gamma \cdot dist_{i \to r,d,t}\} \quad (4)$$

Starting from sender city $i$ on day $d$ and at step $t$, we assume downwind intensity of receiver city $r$ follows an exponential decay as a function of three components (U.S. EPA, 2018; Phillips et al., 2021). The first component is the search radius at the step ($rad_t$), which captures general decreases of downwind intensity over steps. The initial radius is 300 km, and we increase the search radius by 20 km at each step to capture both the uncertainty in the streamline computation and the dispersion of air. The second component is the scalar product of the angle between the receiver city and the wind direction originating from the sender's location ($|\theta|_{i \to r,d,t}$), which means we assign higher intensity to receiver cities that sit closer to the exactly-downwind direction of the sender. The third component is simply the distance between the sender and the receiver city ($dist_{i \to r,d,t}$), which captures geographic decay. We assume $Wind_{i \to r,d,t}$ to be zero if $d_{i \to r,d,t} > rad_t$ (i.e., if receiver city lies outside of the search radius at step $t$) or if $\theta_{i \to r,d,t} > 0.4\pi$ (i.e., if the receiver city is not obviously in the downwind direction from the sender city.[11] Starting from each particular sender city and day, we iterate the procedure for seven steps (i.e., a week).

Perhaps an easier way to think about the procedure above is that, we essentially "move" the location of the sender city with the wind flow and see which receiver cities it touches along the path over the next week period. A visualization of the procedure is shown in Figure 3b. The red arrow at the center represents the locus of the wind flow starting from the sender city. The growing ball of uncertainty around the arrow shows expanding search radius rad over steps in equation (4). The visualization also explains why both the relative angle and distance variables (θ and dist) have starting day and step subscripts (d and t): we track where winds originate and where they move to, and we compute each receiver city's relative angle and distances dynamically.

---

[11] We use parameter values $\{\alpha, \beta, \gamma\} = \{0.7, 0.5, 0.2\}$. These numbers are empirically determined such that we would obtain a spatially continuous flow coefficient function through the successive steps, and that the directionality of the observed winds are respected through the flow coefficient representation. See Online Appendix for more details.

We aggregate step-wise downwind intensity scores to the day level:

$$\text{Wind}_{i \to r,d} = \sum_t \text{Wind}_{i \to r,d,t} \quad (5)$$

and in the econometric analysis below, we further average this city pair-daily score to the monthly frequency ($\text{Wind}_{i \to r,m,y}$ where the variable is indexed by month m, and year y) to make the size of the regression dataset manageable. Because the numerical values of these scores lack direct contextual meaning, in econometric analysis we use a nonparametric approach, grouping this variable into decile bins for positive values plus a separate bin for when downwind score is zero (i.e., because the receiver city is too far away or it is not in the downwind direction, as discussed earlier).

Recall that the goal of this exercise is to predict the atmospheric influence gradient of deforestation. How well does the downwind index—which is built on an arguably parsimonious model—perform in this vein? To test this, we examine how well our index captures sender-to-receiver air pollution passthrough. Our focus is on fine particulate matter pollution (PM2.5), which is known for its long-range transport. We test how the elasticity between PM2.5 pollution in the sender city and the receiver city—data not used in constructing the area-of-effect model—varies as a function of the downwind score.

Appendix Figure 7 reports an estimation of the passthrough between $\text{Pollution}_{i,m,y}$ and $\text{Pollution}_{r,m,y}$ – monthly particulate matter pollution levels at the sender city and at the receiver city – as a function of bins of $\text{Wind}_{i \to r,m,y}$. We estimate the following regression equation:

$$\text{Pollution}_{r,m,y} = \beta \cdot \text{Wind}_{i \to r,m,y} \times \text{Pollution}_{i,m,y}$$
$$+ \gamma \cdot \text{Wind}_{i \to r,m,y} + \delta \cdot \text{Pollution}_{i,m,y} + \alpha_{i,r,m} + \alpha_y + \varepsilon_{i,r,m,y} \quad (6)$$

where $\text{Wind}_{i \to r,m,y}$ enters the regression as decile bins, with the 10th bin representing the weakest wind condition being the reference category and thus omitted from the regression. Equation (6) includes sender-by-receiver-by-month-of-year fixed effects ($\alpha_{i,r,m}$), capturing the identifying variation from within sender-receiver-seasonal pairs across different years, where variations in wind intensities are plausibly exogenous. The pattern of the $\beta$'s in Appendix Figure 7 suggests that indeed the receiver city's measured air pollution concentration respond most strongly to sender city's pollution when the modelled wind transport from the sender to the receiver city is stronger.

## 5.2 Research Design

Our goal is to estimate the relationship between a sender city's quantity of forests and its downwind receiver's environmental and health outcomes, leveraging month-to-month fluctuations in the downwind score (as illustrated in Figure 3b). Our estimation equation is as follows:

$$Y_{r,m,y} = \beta \cdot \text{Forest}_{i,y} \times \text{Wind}_{i \to r,m,y} + \text{Forest}_{i,y} + \text{Wind}_{i \to r,m,y} + \alpha_{i,r,m} + \alpha_y + \varepsilon_{i,r,m,y} \quad (7)$$

A unique observation in our estimation data is indexed by sender city ($i$), receiver city ($r$), month ($m$), and year ($y$). The three key variables in this estimation equation are: $Y_{r,m,y}$, which is the receiver city's outcome in a given month; $\text{Forest}_{i,y}$, which is the sender city's forest acreage in the corresponding year (standardized to mean zero and standard deviation one) multiplied by minus one to capture the effects of losses; and $\text{Wind}_{i \to r,m,y}$, which is the downwind score reflecting strength of wind blowing from $i$ to $r$ in that month. In practice, we categorize the $\text{Wind}_{i \to r,m,y}$ variable into 11 groups: 10 decile bins for non-zero downwind scores and one additional bin for zero wind scores representing "calm" conditions with no downwind relationship between the two cities. We then include $\text{Wind}_{i \to r,m,y}$ as a categorical variable in the regression, and so there is one $\beta$ coefficient for each of the 11 downwind score bins. We include sender by receiver by month-of-year fixed effects ($\alpha_{i,r,m}$), which isolate downwind variation arising within a city pair *and* month of the year to address potential cross-sectional and seasonal confounds. For example, consider the city pair of Sao Paulo and Brasilia. Our specification compares a particular January when the wind blowing from Sao Paulo to Brasilia is particularly strong with other Januaries in different years when the wind strength for the same city pair is comparatively weaker. This strategy enables us to parse out cross-sectional correlations (e.g., the fact that downwind relationships tend to be generally stronger for city pairs where the two cities are in close proximity) and city pair-specific seasonality of wind patterns. We also include year fixed effects ($\alpha_y$) to account for common shocks at the year level. In all regressions, we two-way cluster standard errors at the sender and receiver levels.

Our coefficients of interest are the $\beta$'s, which capture the relationship between upwind forest losses and downwind environmental or health condition at varying levels of wind intensity blowing from the upwind to the downwind city. We hypothesize that at high levels of wind intensity, the coefficient should be positive, indicating that forest losses in the upwind city increases pollution and mortality in the downwind city. As wind intensity decreases, we expect this relationship to weaken. The coefficient estimated under "calm" conditions (i.e., no wind between the city pair) serves as a placebo test: without wind, we anticipate no relationship between upwind forest cover and downwind pollution or health outcomes.

One potential alternative explanation for the observed pollution and mortality effects is a fire channel, where burning in the upwind area—a common practice for deforestation and conversion to farmland—could generate pollution that drifts downwind, thereby reducing air quality and leading to a negative relationship between upwind forest volume and downwind air pollution levels. To assess the relevance of the fire channel, we use remote-sensing data to measure fire activity in upwind cities and then augment equation (7) by adding a full interaction of upwind fire activity with downwind intensity bins. We show below that our forest effect estimates remain largely unchanged after conditioning on the fire effects.

## 5.3 Results

Figure 4 plots the $\beta$ coefficients from equation (7). Each chart represents results for a separate outcome. Start with panel (a), which shows atmospheric outcomes. To facilitate interpretation and comparison of results, we use standardized outcomes (mean 0, standard deviation 1) in each regression, so that the interpretation of the $\beta$ coefficients is "the number of standard deviation change in receiver city's atmospheric outcome per 1 standard deviation decrease in sender city's forest coverage."

The left panel of Figure 4a shows that forests have a large effect on downwind air pollution reduction. The middle and right panels of Figure 4a suggest there is little/null effect of upwind forests on downwind temperatures and precipitation. Two notable patterns are worth pointing out about the pollution effect. First, the forest-pollution effect increases monotonically with the downwind intensity score. In the strongest wind decile bin (bin "1st" on the x-axis), a standard deviation decrease in upwind forests increases downwind pollution concentrations by 3 standard deviations.[12] Second, we find precisely estimated zero effect for the no-wind bin (bin "calm" on the x-axis). This serves as a reassuring "placebo" exercise, suggesting that our econometric specification indicates an absence of forests' downstream environmental effect when our wind analysis indeed suggests an absence of a significant downwind relationship between the city pair.

The empirical impacts of forests on the atmospheric conditions is a subject of ongoing research. Here we compare our findings to some existing studies. On temperature effects, most empirical studies we are aware of document the *local* cooling effects of forests through shading and evapotranspiration (e.g., Han et al., 2021), and so it is not surprising we are not finding a transported effect at large-distance scale. On precipitation effects, our null effects seem at odds with recent studies that show positive impacts of forests on downwind rainfalls (Grosset, Papp, and Taylor, 2023). One important difference of our estimation and Grosset, Papp, and Taylor (2023) is the time scale: while we look at monthly frequency impacts, their findings pertain mostly to longer-term changes. In Appendix Figure 8, we show that, when aggregating our analysis to the annual level, we do see a substantial positive impact of forests on downwind precipitation.[13] Our pollution results echo Xing et al. (2023) which uses a similar design – though looking at within-city scale – to examine the impact of urban forests on city air quality improvement.

We report health effect estimates in Figure 4b. Given the findings of forests' impact on air quality, we separately examine mortality due to cardiovascular/respiratory causes – which are expected to be more responsive to pollution changes – and other, non-cardiorespiratory causes. The left panel of Figure 4b shows that upwind forest losses lead to significant increase in downwind cardiorespiratory mortality rates. Once again, as we saw in the pollution analysis results, the mortality effects are monotonic with respect to downwind intensity score, and we find a precise-zero effect when there are no significant downwind connections between the city pair. The middle panel reports that we find little impact of upwind forests on mortality of non-cardiorespiratory causes. Finally, as an additional "placebo" exercise, in the right panel we look specifically at mortality due to external causes ("accidents") and we find no relationship between upwind forests and external mortality, regardless of downwind intensity levels.

---

[12] We will discuss plausibility of this effect size in conjunction with the health effects estimates.
[13] While forests can contribute to increased local humidity through evapotranspiration, this increased moisture might not be sufficient to alter weather patterns significantly in the downwind area, especially over short distances or periods. Over the long term, extensive forests can modify regional climate patterns. As forests mature, they contribute to higher levels of evapotranspiration. This increased moisture in the air can eventually influence weather patterns, potentially leading to more rainfall in downwind areas.

In Appendix Figure 10, we present a heterogeneity analysis in which we use land cover transition information to classify each upwind city-year's forest volume based on its land cover type from the previous year, distinguishing how much of the current year's forest acreage was previously forest, non-forest vegetation land, agricultural land, or non-vegetated land (i.e., urbanized area). We then separately estimate the impacts of upwind forests on downwind air quality and cardio-respiratory mortality based on the previous year's land cover. We found that environmental and health benefits primarily arise when upwind areas were forested in the previous year, as opposed to other land cover types.

In Appendix Figure 11, we report our main downwind pollution and mortality estimates after conditioning on the impact of upwind fire activities. The results remain virtually unchanged, indicating that fire-related pollution is unlikely to be driving the forest effects we observe.

The magnitude of our pollution and health effects estimates are generally in line with prior literature on the causal effects of air pollution. Our pollution regression in Figure 4a suggests that, with the strongest downwind score, each standard deviation decrease in forests increases pollutants by 3 standard deviations, which would convert to about 65 ug/m3 increase in $PM_{10}$. The mortality regression in Figure 4b suggests the corresponding change in cardiorespiratory mortality rate is about 0.37 deaths per 100,000 people. These numbers imply about a $PM_{10}$-mortality elasticity of 0.99%. In comparison, in one of the most cited quasi-experimental studies in the U.S., the estimated mortality-$PM_{2.5}$ elasticity among the elderly population aged 65 and above is 1.86% (Deryugina et al., 2019).

## 6. Discussion and Conclusion

We are now ready to draw conclusions by plugging in various pieces of our empirical estimates back into the conceptual equation (1) we laid out in Section 3. The empirical version of the equation is as follows:

Excess Deaths due to city $i$'s agricultural trade

$$= \sum_{r,m,y} (\beta_{\Delta Trade \to \Delta Forest} \cdot \Delta Trade_i) \cdot (\beta^{i \to r,m,y}_{Forest \to CR\ mortality} \cdot Population_{r,y}) \quad (8)$$

That is, the total excess deaths due to city $i$ equals the product of trade-induced deforestation in $i$ (the first parenthesis) and excess deaths per unit of deforestation at a receiver city $r$ (the second parenthesis), summed across all receiver cities and time periods. The subscript of the term $\beta^{i \to r,m,y}_{Forest \to CR\ mortality}$ means we only consider the impact of forest on downwind cardiovascular mortality – what we found to be responding to upwind forest changes; the superscript indicates we use nonlinear estimates as shown in Figure 4, panel (b) based on the downwind index associated with the city pair $i \to r$, month $m$, and year $y$.

One key qualitative insight from this calculation is that trade shocks (and the resulting deforestation) occur in areas distinct from where the associated mortality burden is felt. Figure 5, panel (a) shows a map of trade-induced deforestation based on our estimates. In panel (b), we plot the distribution of total mortality burdens at receiver cities. That is, for each city, we sum up excess deaths due to deforestation from all upwind cities over the study period. The difference of the spatial patterns between the two panels is apparent: because of wind directivity and population distribution, areas with large numbers of excess deaths may situate hundreds if not thousands of miles away from where trade deforestation is happening.

Quantitatively, our estimates suggest a total of 3.6 million ha loss of Brazilian forests due to trade, resulting in over 732,000 excess deaths over the study period. For reference, total deaths in Brazil in the year of 2018 is estimated to be around 1.28 million. We quantify the mortality impacts in terms of its statistical life value, which is a common way to think about the costs (values) of environmental damage (protection) (e.g., U.S. EPA, 2000). Because direct estimation of the value of a statistical life (VSL) for Brazil is lacking, we use a transfer approach that downscales the U.S.-based VSL estimate of $2.3 million (Ashenfelter and Greenstone, 2004) by factors of a transfer elasticity of 1.2 (Narain and Sall, 2016) times a Brazil-U.S. income per capita

ratio of 7. This yields a VSL estimate of $0.7 million USD. The total statistical life value loss of the 732,000 extra deaths thus amounts to about 513 billion USD. This represents a significant number – about 18 percent of the total agricultural export value of Brazil over the study period.

Our estimated elasticity of 0.18 may underestimate the actual health effects of trade-induced deforestation, as it does not account for morbidity, long-term health impacts beyond same-month pollution exposure, or productivity and income losses that could indirectly worsen health outcomes. In addition to air pollution, trade-induced deforestation may also degrade water quality, alter microclimates, and disrupt ecosystems, all of which could further affect health and increase the actual elasticity between agricultural exports and mortality.

On the policy front, our study does not suggest that agricultural trade is inherently negative. Trade can generate multiplier effects in the exporting country through sectoral changes, employment, and migration, leading to welfare gains that exceed the value of the exports themselves. However, our paper highlights a large negative aspect of trade on health and the resulting regional inequality, as the mortality costs and income benefits may not be distributed in the same areas.

With increased globalization, to meet foreign demand without high health costs, agricultural producers should prioritize increasing yield rather than expanding agricultural lands. This requires domestic policies for deforestation monitoring, which are shown to be effective in our heterogeneity results across presidential administrations. International pressure or green trade policy can also effectively avoid deforestation shown by existing studies (e.g. Nolte et al., 2013; Kerr, 2013; Hsiao, 2021).

For future research, our paper provides a framework to estimate the value of forest and other natural capitals. Our identification strategy relies on foreign demand changes that serve as exogenous drivers of forest loss, as well as quasi-random wind flow that separates upwind drivers and downwind recipients. For the latter, future studies may exploit other networks that connect spatially different regions. In addition, our wind trajectory simulation is also helpful for future environmental economics studies to estimate the impacts of air pollution.

**Figure 1. Trends in Trade, Agricultural Expansion and Deforestation**

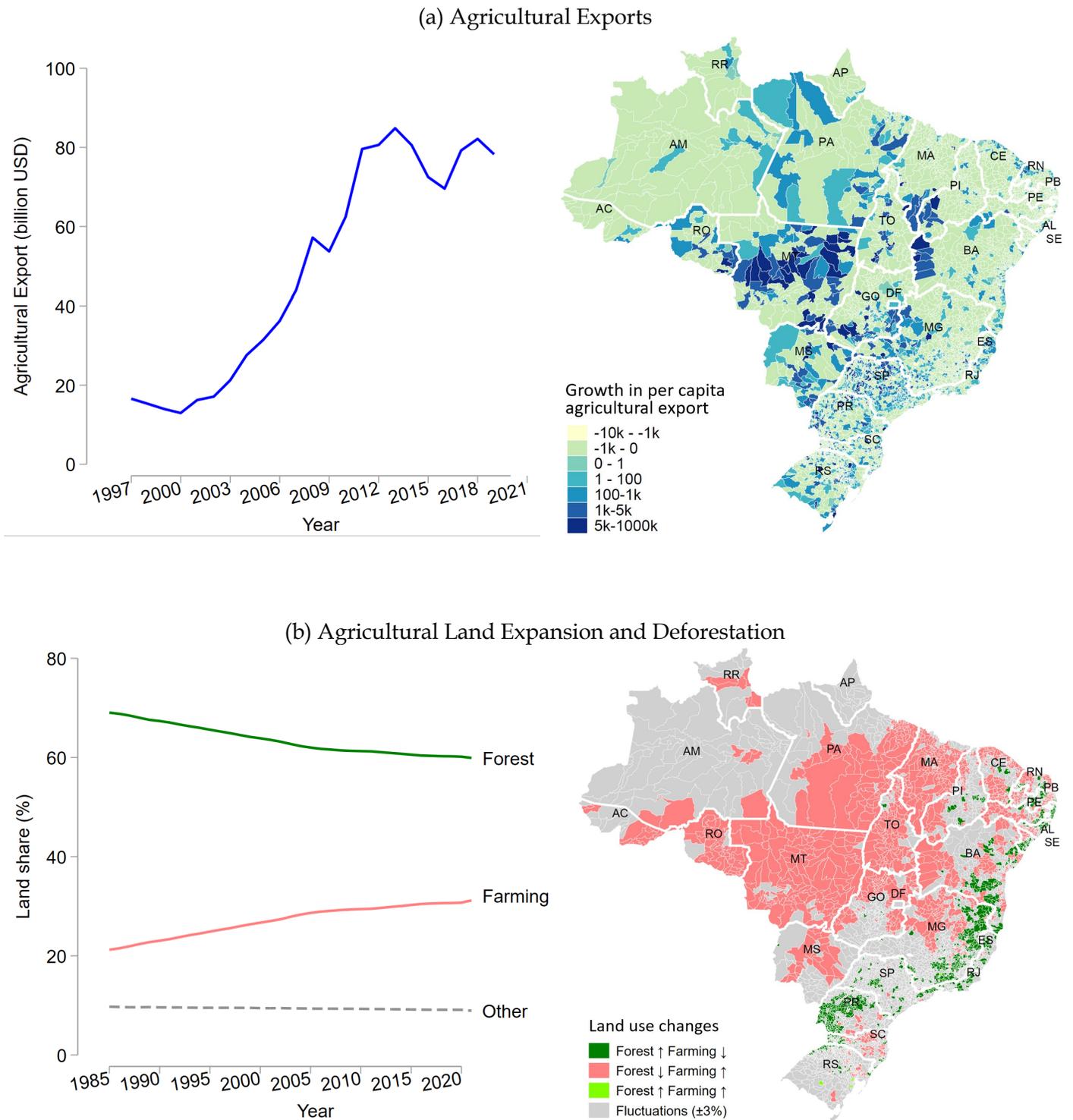

*Notes:* Panel (a) shows temporal and geographic trends in Brazil's agricultural exports. Panel (b) shows temporal and geographic trends in the substitution between agricultural land and forests. "Fluctuations" correspond to areas with less than 3 percent changes in forest or farmland coverage over the study period.

**Figure 2. The Effect of Trade on Deforestation and Agricultural Expansion**

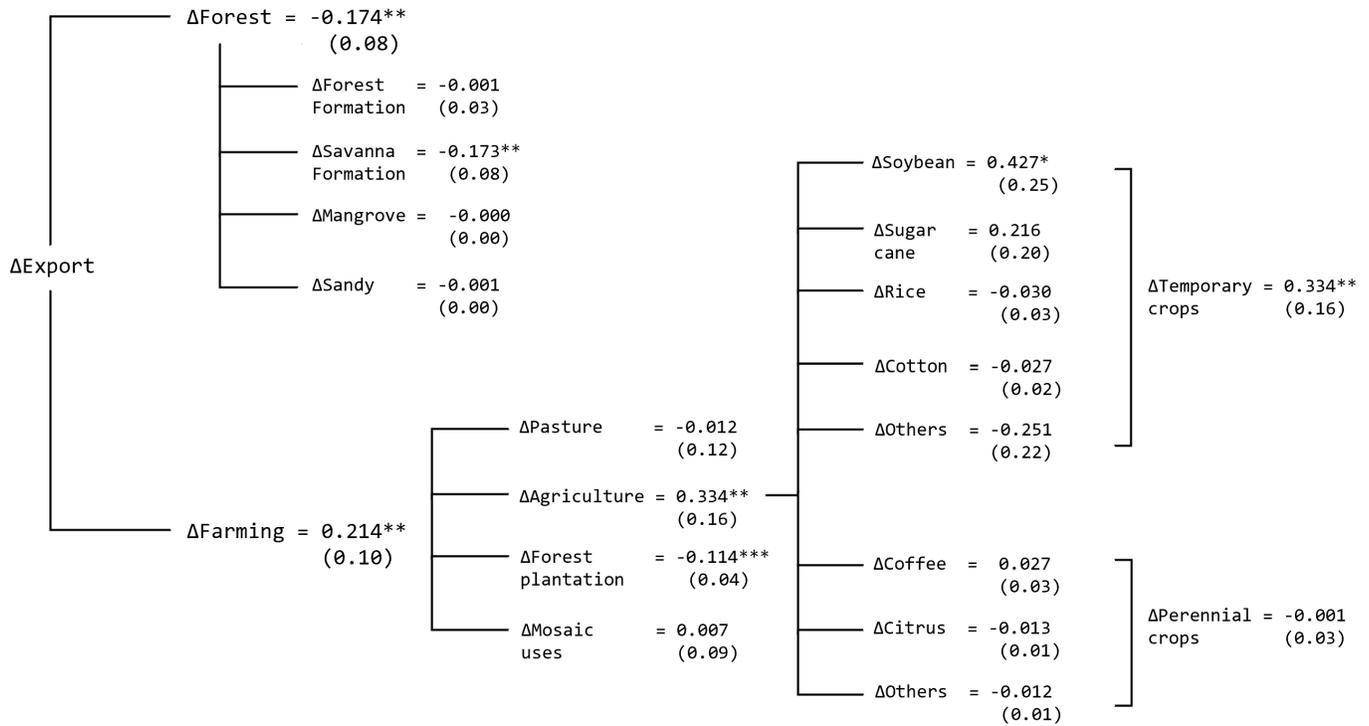

*Notes:* Each node on the tree represents a separate IV regression following the exact same specification except for the outcome variable, which is denoted by the name of the node. Each regression has 57,189 underlying observations with a first-stage Kleibergen-Paap F-statistic of 8.1. Branches of the tree represent hierarchies of the land use categorizations. Interpretation of the coefficients is change in land cover land use type (in percentage points) per 1,000 BRL increase in export per capita.

**Figure 3. Area-of-Effect Modeling**

(a) Airflow Modeling

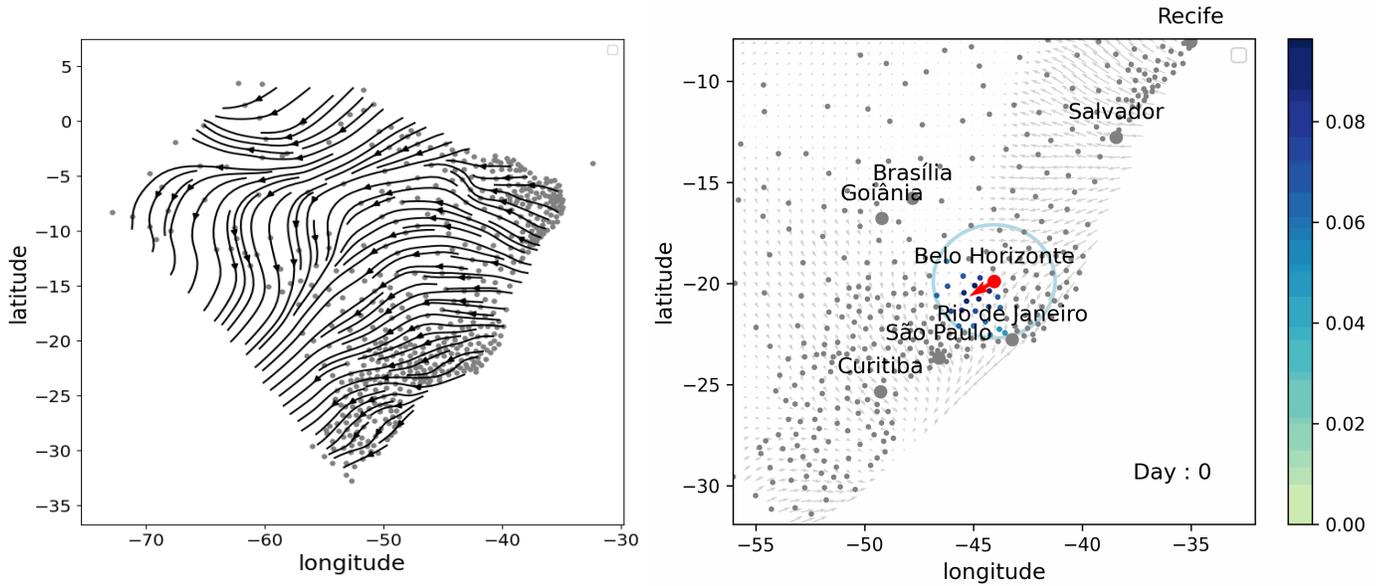

(b) Downwind Score: Belo Horizonte as Example

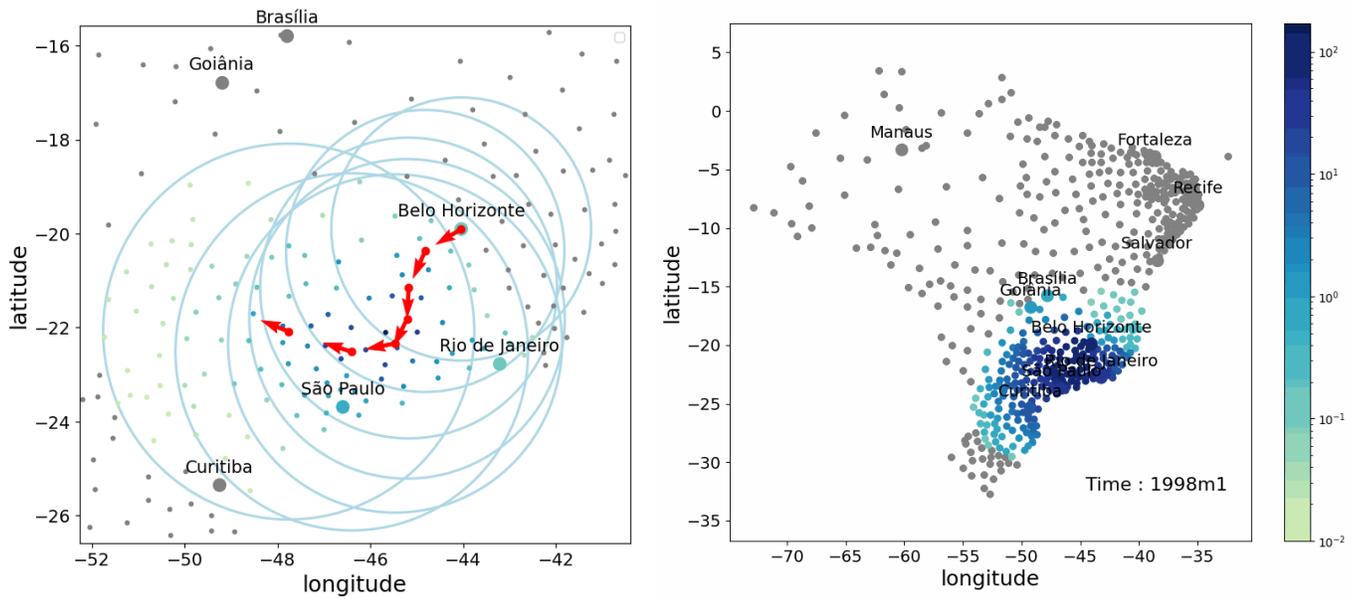

*Notes:* Figures show dotted map of Brazil where each dot represents centroid of a city. Panel (a) shows a day snapshot of wind flows built from location-specific wind vectors data (left) and a visualization of our wind flow model that tracks downwind cities influenced by winds from a sender city (Belo Horizonte in this example) over the course of seven steps (i.e., days). Panel (b) shows aggregation of step-specific downwind scores to sender-receiver-day level (left) and sender-receiver-month level (right) where the sender city is Belo Horizonte.

**Figure 4. The Downwind Effects of Forest Losses**

(a) Atmospheric Outcomes

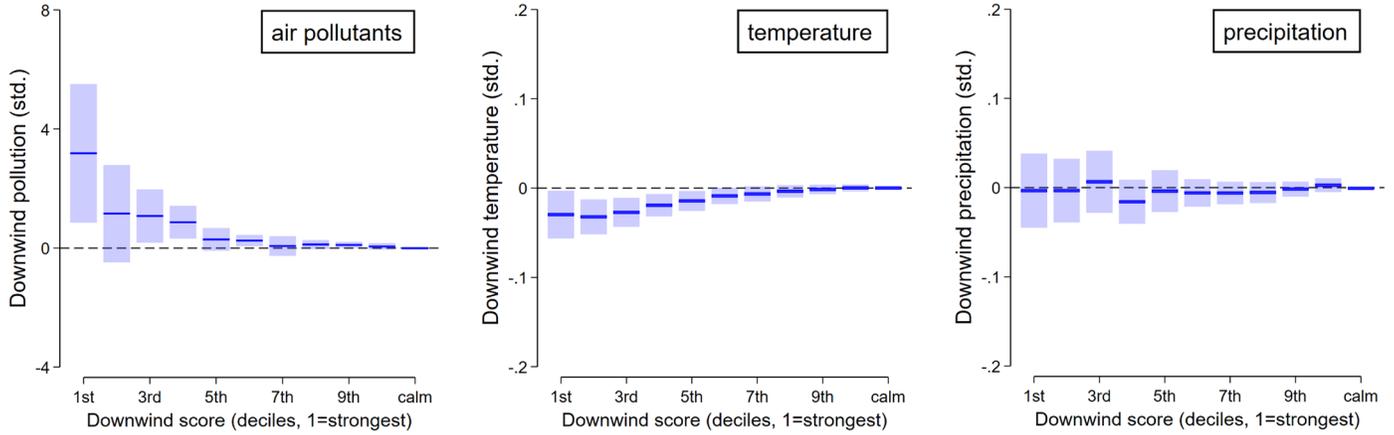

(b) Mortality Outcomes

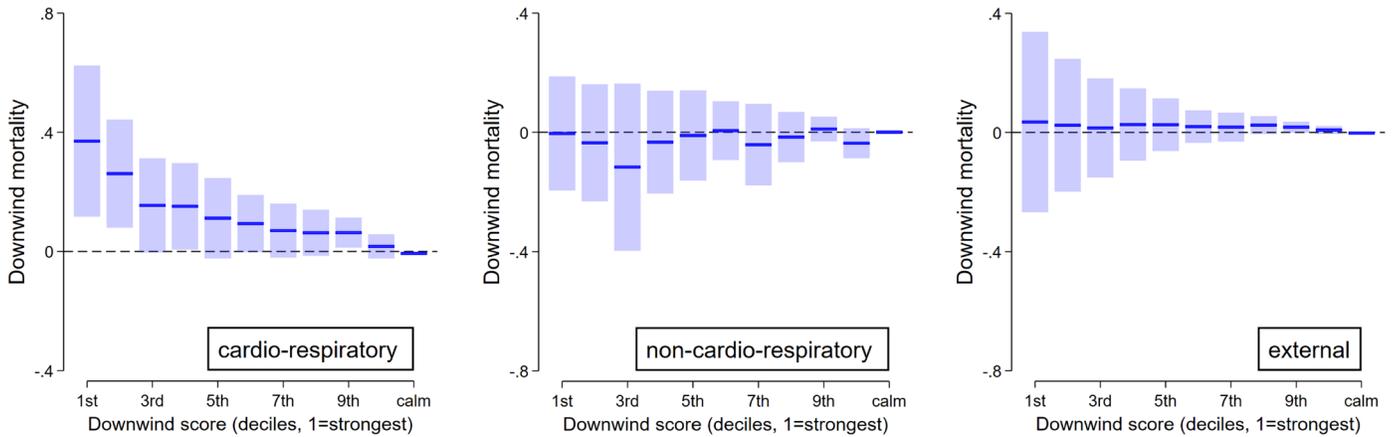

*Notes:* Charts show estimates on changes in downwind outcomes per 1 SD decrease in upwind forest cover, separately by downwind exposure score bins. Each chart shows a separate regression following the exact same specification except for the outcome variable. Within each chart, horizontal step lines show point estimates, and range bars show 95 percent confidence intervals.

**Figure 5. Distribution of Trade-Induced Forest Losses and Excess Deaths**

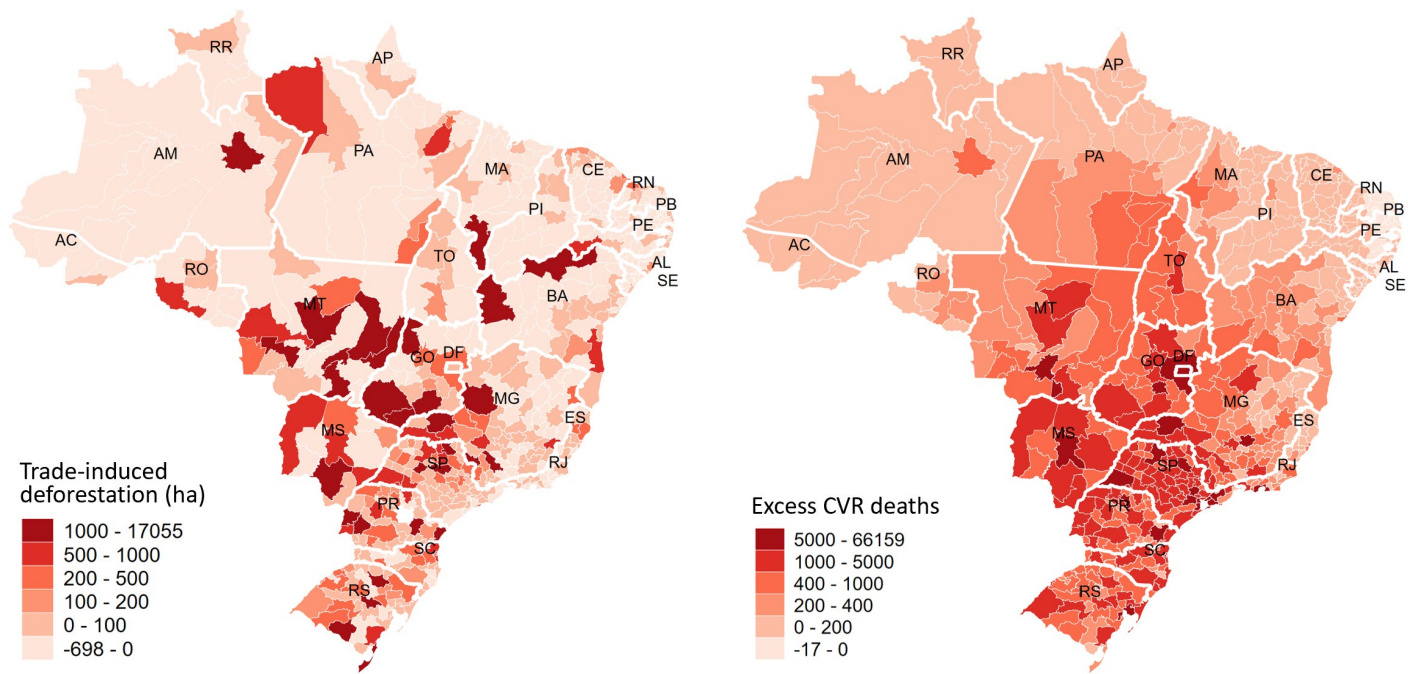

*Notes:* The map on the left shows the spatial distribution of trade-induced deforestation over the study period. The map on the right shows the spatial distribution of the downstream cardio-respiratory deaths due to trade-induced deforestation.

**Appendix Figures and Tables**

**Appendix Figure 1. Brazil's Agricultural Export Structure**

(a) Product Categories

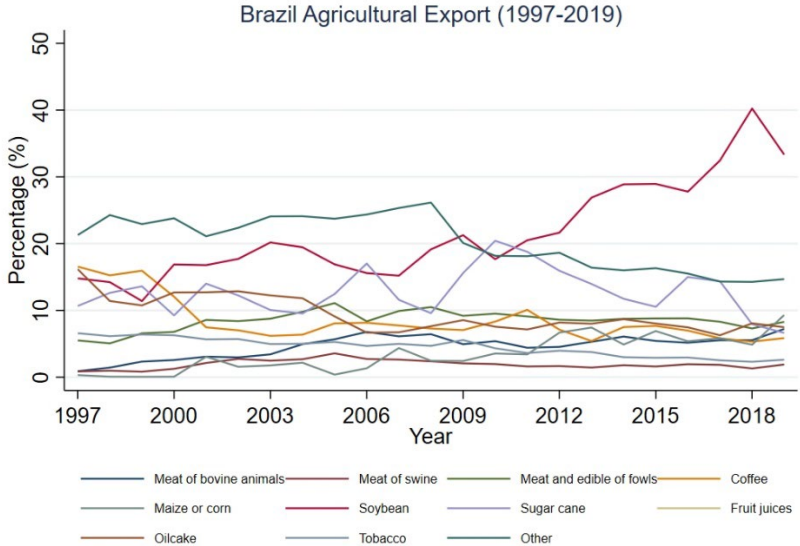

(b) Trade Partners

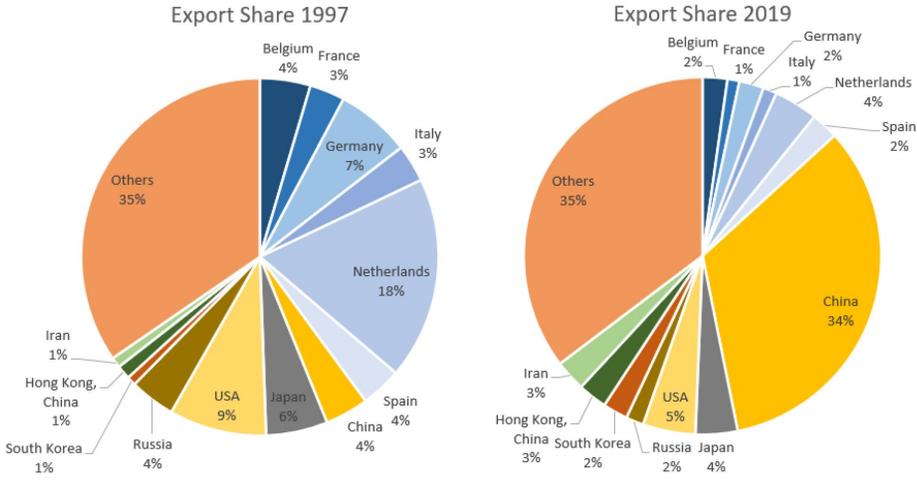

*Notes:* Panel (a) shows distribution of export values in billions of USD by product category. Panel (b) shows agricultural export share by trade partner in 1997 (left) and in 2019 (right).

**Appendix Figure 2. Geography of Land Use Changes**

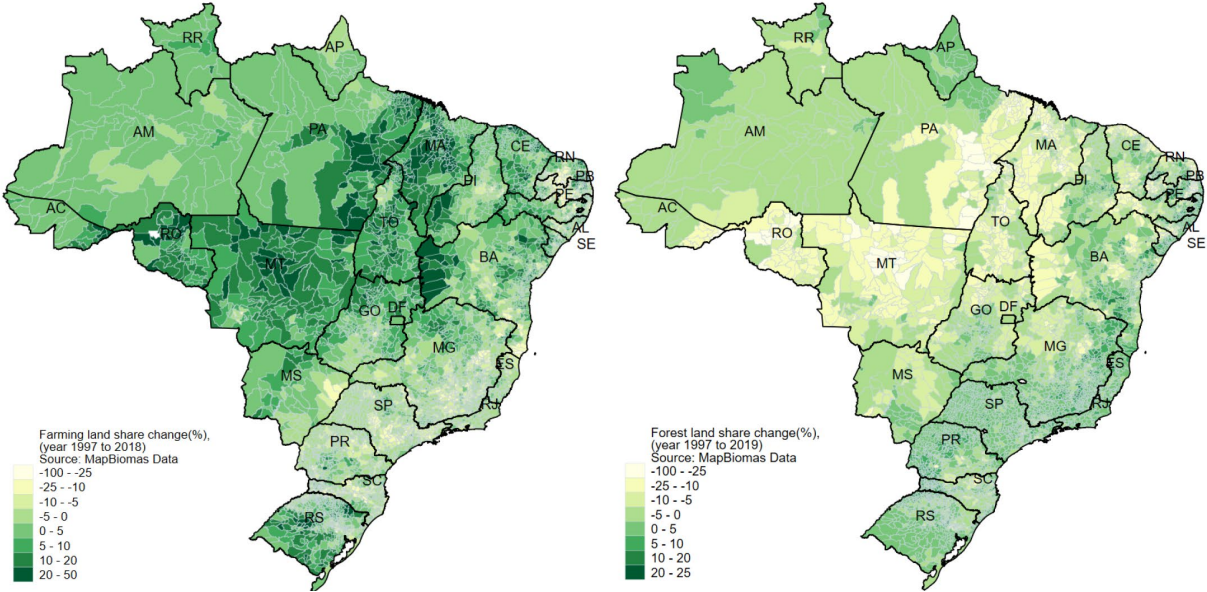

*Notes:* Maps show 1997-2018 percentage change in land use for farming purposes (left) and forest land (right).

## Appendix Figure 3. Geography of Agricultural Growth

(a) Agricultural Employment Growth

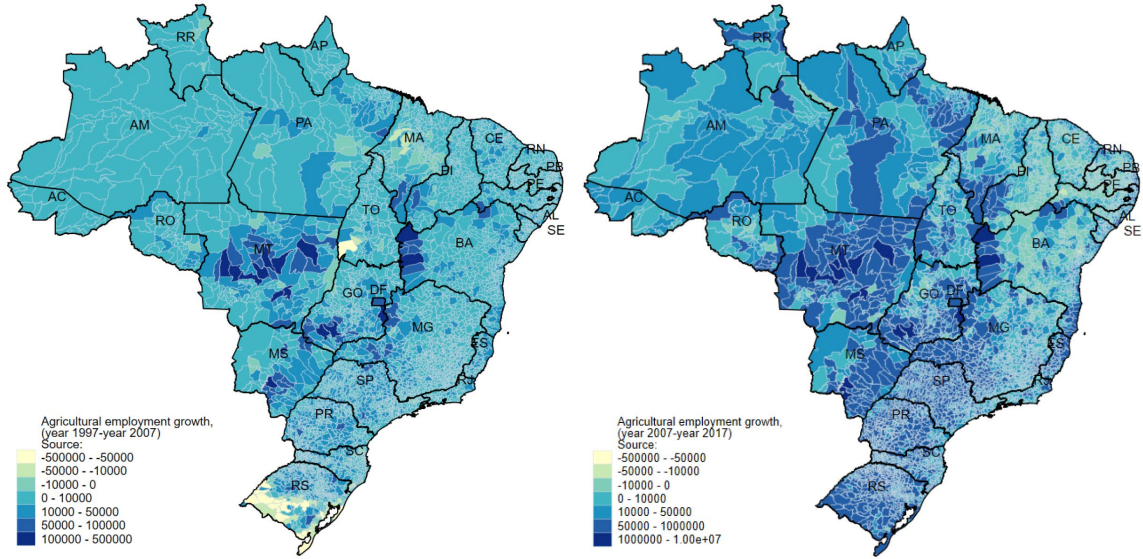

(b) Agricultural Export Growth

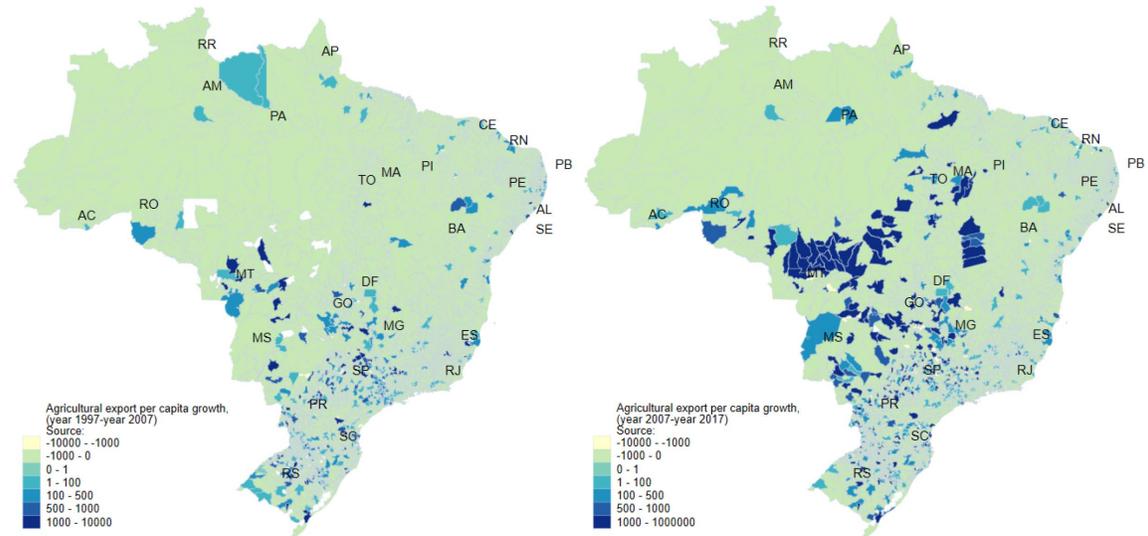

*Notes:* Panel (a) shows growth in agricultural employment over the 1997-2007 period (left) and the 2007-2019 period (right). Panel (b) shows growth in agricultural exports per capita over the 1997-2007 period (left) and the 2007-2019 period (right).

# Appendix Figure 4. Geography of Mining Activities

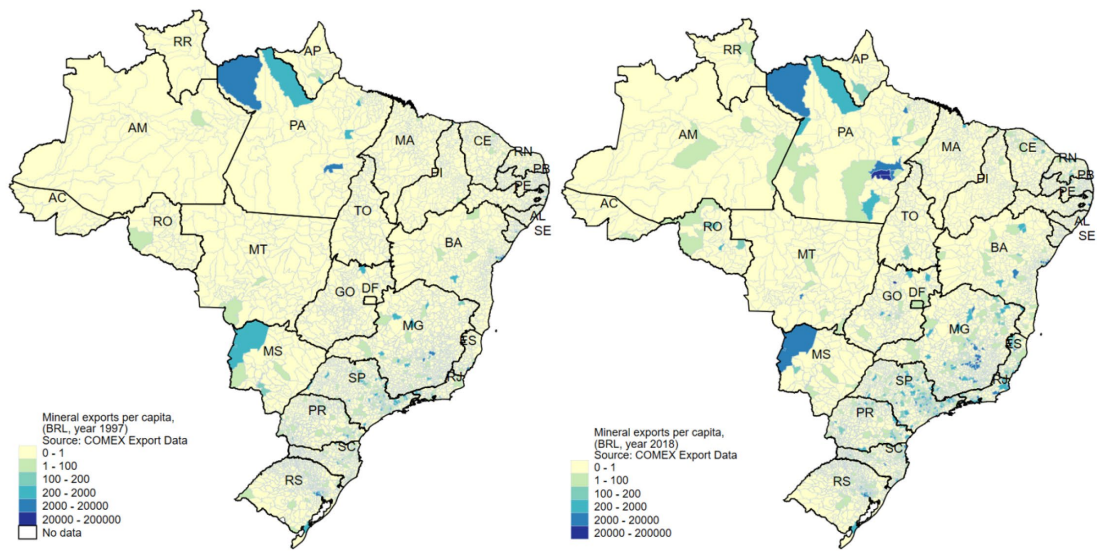

*Notes:* Maps show mineral exports per capita in year 1997 (left) and in year 2018 (right).

# Appendix Figure 5. Export Demand Shock Variability

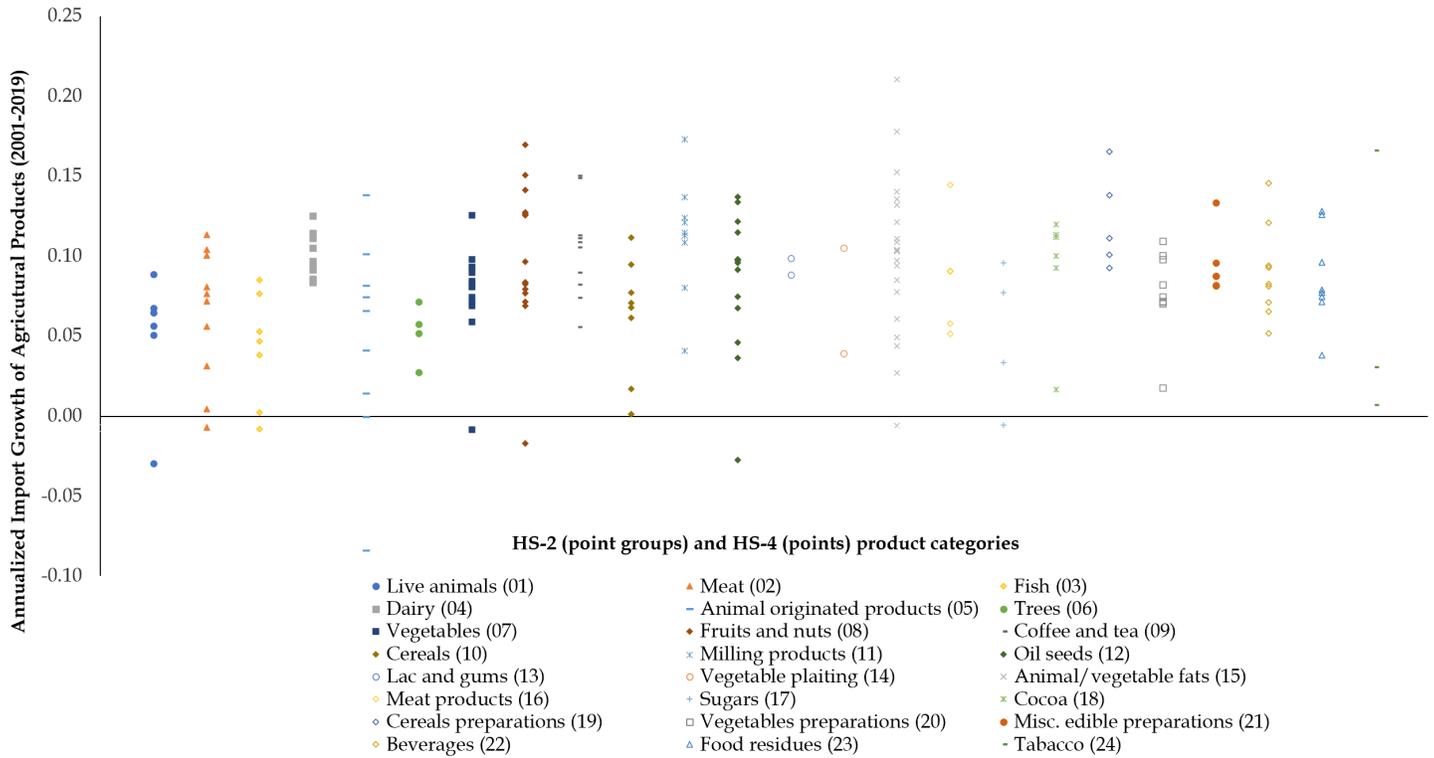

*Notes:* Major importing countries' export demand growth rates variation across HS-4 product categories, represented by points. Points are grouped into broader HS-2 categories.

**Appendix Figure 6. Effects of Placebo Export Shocks on Forest Cover**

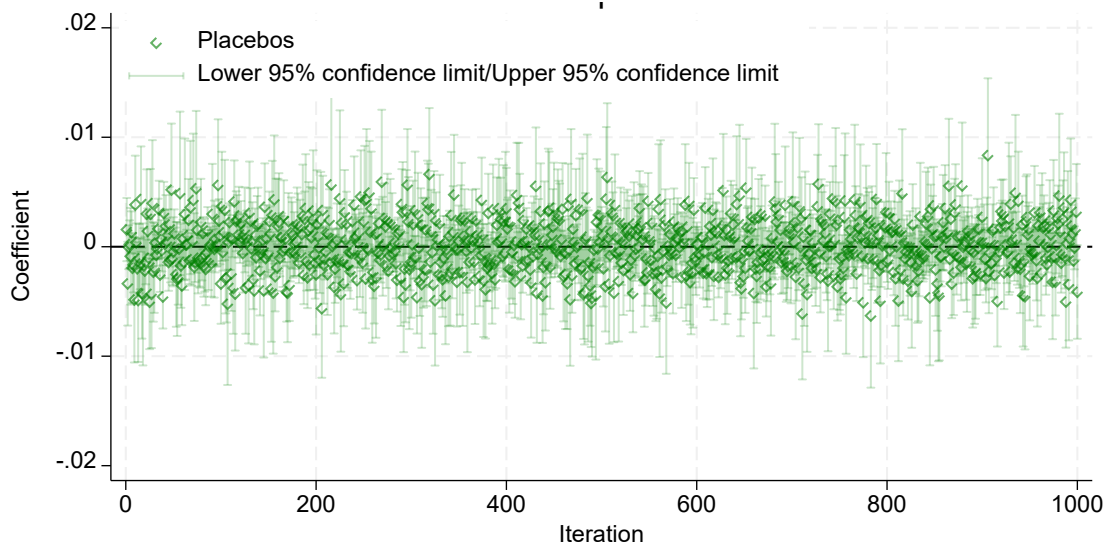

*Notes:* This figure plots the coefficient on placebo shocks in 1,000 separate regressions, where the dependent variable is the forest land use 4 years growth. The placebo shock is the placebo change in agricultural import growth by HS code and municipality, a normally distributed random variable with mean 0 and variance 5. The regression also contains year and macroregion fixed effects, and agricultural employees are used as weights. Standard errors are clustered at the AMC level.

**Appendix Figure 7. Passthrough of Air Pollution from Upwind to Downwind Cities by Downwind Score Deciles**

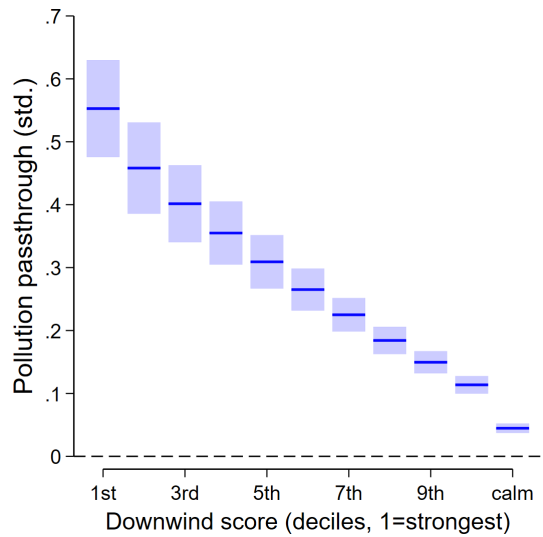

*Notes:* This figure shows coefficients from a regression of a receiver city's PM2.5 concentration on an upwind city's PM2.5 concentration, with the effect allowed to vary by the downwind score from the upwind location to the downwind location according to our area-of-effect model. All regression controls for city pair by month-of-sample fixed effects and year fixed effects. Standard errors are two-way clustered at the sender city and receiver city levels. Range bars show 95 percent confidence intervals.

**Appendix Figure 8. The Downwind Effect of Forest Losses on Precipitation: Month-Of versus Annual Effects**

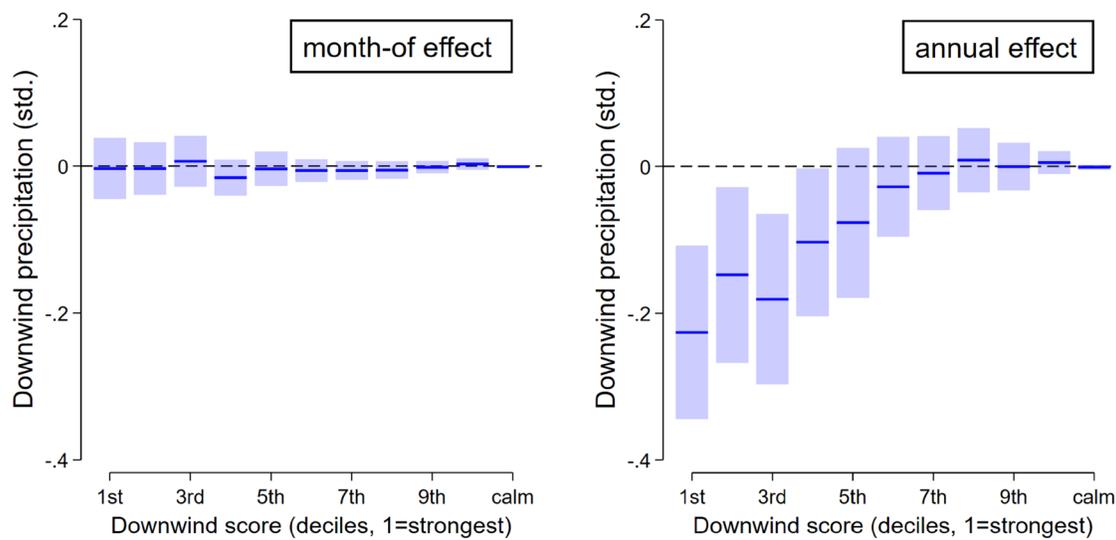

*Notes:* Charts show estimates on changes in downwind precipitation outcomes per 1 SD decrease in upwind forest cover. Left panel shows regression estimates based on monthly data. Right panel shows regression estimates from the same data aggregated to the annual level. The annual regression equation, analogous to the monthly equation (7), controls for sender-by-receiver fixed effects and year fixed effects, with standard errors two-way clustered at the sender and receiver levels.

**Appendix Figure 9. The Downwind Effect of Forest Losses on Air Pollutants**

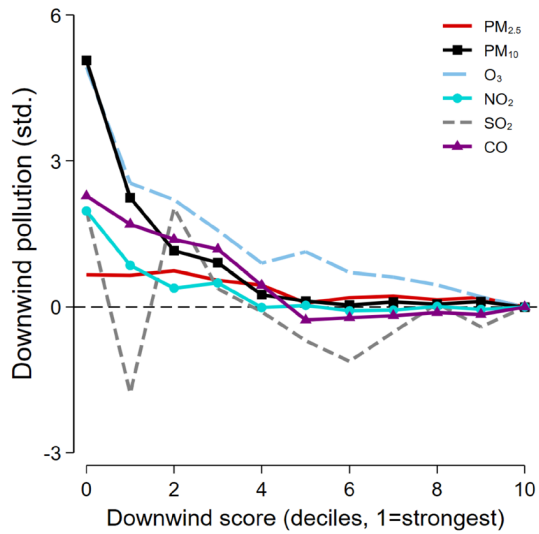

*Notes:* Chart shows estimates on changes in downwind pollution outcomes per 1 SD decrease in upwind forest cover. Each line represents a separate regression using standardized pollutant concentrations as the outcome variable.

**Appendix Figure 10. The Downwind Effect of Forests: Heterogeneity by Prior Land Type**

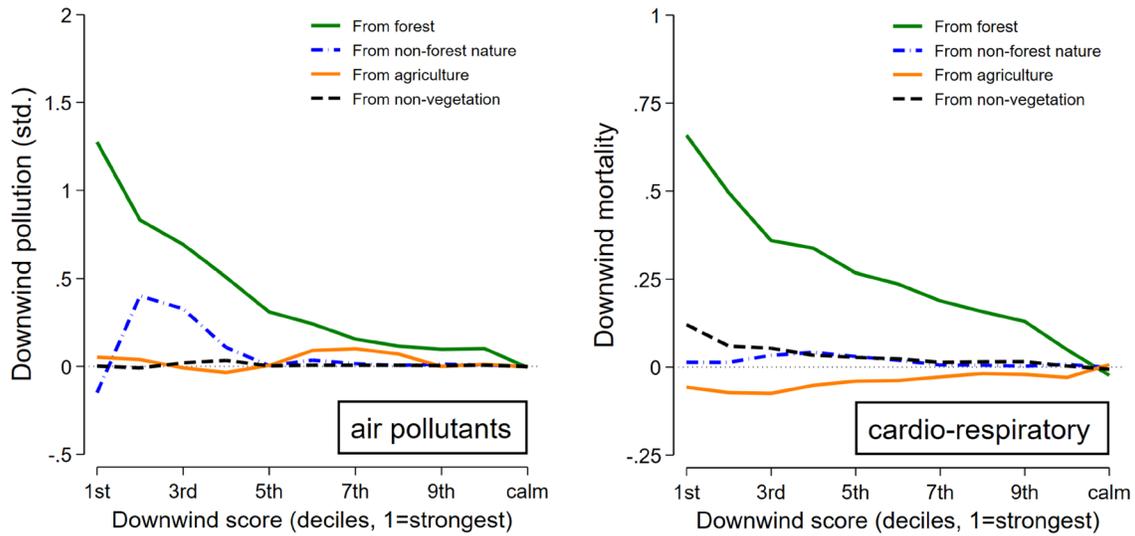

*Notes:* Charts show estimates on changes in downwind outcomes per 1 SD decrease in upwind forest cover, separately by land cover in the previous year. We use land cover transition information to classify each upwind city-year's forest volume based on its land cover type from the previous year, distinguishing how much of the current year's forest acreage was previously forest, non-forest vegetation land, agricultural land, or non-vegetated land (i.e., urbanized area). We then separately estimate the impacts of upwind forests losses on downwind air quality and cardio-respiratory mortality based on the previous year's land cover.

**Appendix Figure 11. The Downwind Effects of Forest Losses: Fire Controls**

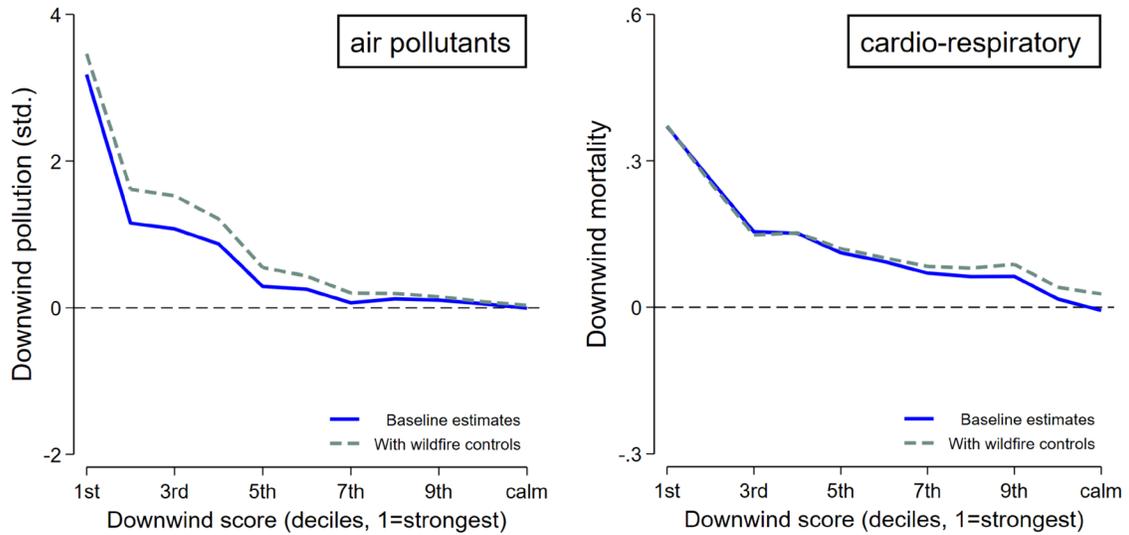

*Notes:* Charts show estimates on changes in downwind outcomes per 1 SD increase in upwind forest cover, separately by downwind exposure score bins. Each chart shows two separate regression, one with and the other without controls for upwind fire occurrences.

# Appendix Table 1. Major Import Countries Selection

| Country | Rank in 2011 | Years with missing UN total import data | Years with missing UN import to Brazil and missing Comex export data | Years with low correlation in 2010-2019: UN import to Brazil vs. Comex export (below 0.8) | Mismatch between UN import data HS4 and HS2 (difference>3%, 23y*24 HS2, <27 mismatches) | Selected |
|---|---|---|---|---|---|---|
| USA | 1 | 0 | 0 | 0 | 0 | Yes |
| Germany | 2 | 0 | 0 | 0 | 200 | No |
| Japan | 3 | 0 | 0 | 0 | 0 | Yes |
| China | 4 | 0 | 0 | 0 | 0 | Yes |
| United Kingdom | 5 | 0 | 0 | 0 | 21 | Yes |
| Netherlands | 6 | 0 | 0 | 0 | 11 | Yes |
| France | 7 | 0 | 0 | 0 | 2 | Yes |
| Italy | 8 | 0 | 0 | 0 | 96 | No |
| Belgium | 9 | 2 | 2 | 3 | 0 | No |
| Russian Federation | 10 | 0 | 0 | 0 | 0 | Yes |
| Spain | 11 | 0 | 0 | 0 | 72 | No |
| Canada | 12 | 0 | 0 | 0 | 0 | Yes |
| Rep. of Korea | 13 | 0 | 0 | 0 | 0 | Yes |
| Mexico | 14 | 0 | 0 | 2 | 87 | No |
| China, Hong Kong SAR | 15 | 0 | 0 | 0 | 0 | Yes |
| Nigeria | 16 | 2 | 2 | 1 | 3 | No |
| Saudi Arabia | 17 | 1 | 1 | 0 | 0 | No |
| Poland | 18 | 0 | 0 | 0 | 0 | Yes |
| Indonesia | 19 | 0 | 0 | 0 | 0 | Yes |
| Malaysia | 20 | 0 | 0 | 0 | 0 | Yes |
| Sweden | 21 | 2 | 0 | 0 | 0 | No |
| India | 22 | 0 | 0 | 0 | 8 | Yes |
| Egypt | 23 | 1 | 0 | 0 | 8 | No |
| United Arab Emirates | 24 | 3 | 23 | 0 | 15 | No |
| Denmark | 25 | 0 | 0 | 1 | 17 | No |
| Austria | 26 | 0 | 0 | 10 | 0 | No |
| Switzerland | 27 | 1 | 0 | 10 | 3 | No |
| Singapore | 28 | 0 | 0 | 0 | 53 | No |
| Australia | 29 | 0 | 0 | 0 | 0 | Yes |
| Portugal | 30 | 0 | 0 | 0 | 4 | Yes |
| Brazil | 31 | 0 | 21 | 2 | 0 | No |
| Other Asia, nes | 32 | 0 | 23 | 0 | 12 | No |
| Thailand | 33 | 1 | 0 | 0 | 1 | No |
| Türkiye | 34 | 1 | 0 | 0 | 23 | No |
| Algeria | 35 | 2 | 2 | 0 | 2 | No |
| Iran | 36 | 5 | 6 | 4 | 2 | No |
| Viet Nam | 37 | 3 | 3 | 0 | 0 | No |
| Greece | 38 | 0 | 0 | 0 | 0 | Yes |
| Czechia | 39 | 0 | 0 | 10 | 39 | No |
| Ireland | 40 | 0 | 0 | 8 | 3 | No |
| Bangladesh | 41 | 6 | 6 | 0 | 11 | No |
| Norway | 42 | 0 | 0 | 10 | 1 | No |
| Venezuela | 43 | 6 | 6 | 0 | 8 | No |
| South Africa | 44 | 3 | 3 | 0 | 2 | No |
| Philippines | 45 | 0 | 0 | 0 | 0 | Yes |
| Ukraine | 46 | 1 | 0 | 2 | 4 | No |
| Romania | 47 | 0 | 0 | 0 | 119 | No |
| Finland | 48 | 0 | 0 | 0 | 10 | Yes |
| Morocco | 49 | 0 | 0 | 0 | 6 | Yes |
| Chile | 50 | 0 | 0 | 0 | 0 | Yes |

*Notes:* This table tabulates Brazil's major export destinations ranked by Brazil's export value, highlighting countries that are selected in constructing the shift-share instruments. Excluded countries are those with either missing UN import or Comex export data, low correlation between UN import and Comex export data, or those with high numbers of mismatches between UN import HS-4 and HS-2 product categories.

**Appendix Table 2. Summary Statistics**

| Variable Name | Obs | Mean | SD |
|---|---|---|---|
| $\Delta$Forest land share (%) | 57189 | -0.13 | 2.494 |
| $\Delta$Farming land share (%) | 57189 | 0.05 | 2.666 |
| $\Delta$Pasture land share (%) | 57189 | -1.10 | 4.194 |
| $\Delta$Agriculture land share (%) | 57189 | 0.98 | 2.754 |
| $\Delta$Forest plantation land share (%) | 57189 | 0.26 | 0.732 |
| $\Delta$Mosaic of uses land share (%) | 57189 | -0.09 | 2.884 |
| $\Delta$Export per capita (1000 BRL) | 57189 | 0.22 | 1.327 |
| $\Delta$Export per capita (real 1000 BRL) | 57189 | 0.19 | 1.562 |
| Population | 57189 | 48431.26 | 2.45e5 |
| Agricultural employee | 57189 | 23721.17 | 1.82e5 |
| Income per capita | 57189 | 191.40 | 114.516 |
| Literacy rate (%) | 57189 | 70.32 | 16.445 |
| Population density | 57189 | 0.82 | 3.808 |
| Rural population | 57189 | 9398.84 | 16462.777 |
| Urban population | 57189 | 28374.83 | 1.99e5 |
| GE soy | 57189 | 0.03 | 0.005 |

*Notes:* The variables for land use and export per capita describe the growth over four years within the period from 2001 to 2019; Real export values have been deflated by the Brazilian Consumer Price Index (IPCA) calculated by IBGE and are denominated in 2019 reals; Income per capita, literacy rate, population density, rural population, and urban population are values from 1991, calculated by IBGE.

## Appendix Table 3. Shift-Share Balancing Tests

|  | (1) Coef. | (2) S.E. | (3) $p$ value | (4) $q$ value |
|---|---|---|---|---|
| Urban population | 4704.262 | 3952.672 | 0.234 | .429 |
| Rural population | 3050.763 | 2608.679 | 0.242 | .429 |
| Total population | 7755.025 | 6395.578 | 0.225 | .429 |
| Per capita income | 4.635 | 1.635 | 0.005 | .06 |
| Literacy rate | 0.264 | 0.151 | 0.081 | .429 |
| Population density | -0.003 | 0.007 | 0.691 | .754 |
| GE soy seeds | 0.000 | 0.000 | 0.608 | .73 |
| Average temperature | 0.030 | 0.031 | 0.332 | .498 |
| Average humidity | -0.005 | 0.078 | 0.947 | .947 |
| Average pressure | -0.478 | 0.416 | 0.250 | .429 |
| Average wind speed | -0.006 | 0.007 | 0.383 | .511 |
| Total precipitation | 652.890 | 438.455 | 0.137 | .429 |

*Notes:* Export values are deflated by the Brazilian Consumer Price Index (IPCA) calculated by IBGE and are denominated in 2019 reais. The number of AMC regions is 3812. All regressions include Year fixed effects and macroregion fixed effects. The independent variable is the four-year shift-share instrument generated for the growth of Brazilian agricultural exports. "*q* value" is the False Discovery Rate adjusted significance level (Anderson, 2008)

**Appendix Table 4. Trade-Deforestation Regression Estimates: Alternative Time Lags**

|  | (1) | (2) | (3) | (4) | (5) | (6) |
|---|---|---|---|---|---|---|
| **Panel A. First stage** | Export (Δ1y) | Export (Δ2y) | Export (Δ3y) | Export (Δ4y) | Export (Δ5y) | Export (Δ6y) |
| Shift-share IV | 0.21 | 1.12*** | 1.78*** | 0.91*** | 1.12*** | 1.04*** |
|  | (0.14) | (0.12) | (0.30) | (0.23) | (0.25) | (0.28) |
| **Panel B. Second stage** | Forest (Δ1y) | Forest (Δ2y) | Forest (Δ3y) | Forest (Δ4y) | Forest (Δ5y) | Forest (Δ6y) |
| ΔExport per capita (1000 BRL) | -0.118 | -0.038** | -0.043* | -0.174** | -0.207*** | -0.192** |
|  | (0.12) | (0.02) | (0.02) | (0.08) | (0.07) | (0.10) |
| Observations | 68,585 | 68,585 | 64,814 | 57,189 | 49,564 | 41,939 |
| First-stage F (Kleibergen-Paap) | 1.250 | 81.63 | 41.01 | 8.072 | 16.59 | 6.492 |
| Year FE | Yes | Yes | Yes | Yes | Yes | Yes |
| Macroregion FE | Yes | Yes | Yes | Yes | Yes | Yes |
| Control | Yes | Yes | Yes | Yes | Yes | Yes |

*Notes:* The IVs are constructed based on the agricultural imports by top import countries. All monetary values have been deflated by the Brazilian Consumer Price Index (IPCA) calculated by IBGE and are denominated in 2019 reals. *: $p < 0.10$; **: $p < 0.05$; ***: $p < 0.01$.

# Technical Appendix: Area of Effect Estimation Details

To build a matrix that summarizes monthly wind flow intensities between all pairs of cities in Brazil, we designed a model that evaluates the intensity of the wind between a sender city and a receiver city at a given date. Beginning from a particular day and city, the model constructs streamlines by sequentially following the wind's speed and direction on a daily basis.

The input wind data (wind direction and speed information, i.e., vectors) are at a city-day level. Thus, the first step is to construct for each day a wind vector field with a grid resolution to estimate wind vectors at any point of the Brazilian territory. For that purpose, we divided the largest dimension of Brazil $dim$ in polar coordinates (its width) by $res = 64$. Then, we built a coordinates grid with a resolution of $dim/res \approx 0.6$. The wind vector grid is then computed using linear interpolation from the city-day level wind data on the coordinates grid. Note that the higher the resolution $res$, the longer the computation so the value of $res$ was chosen to obtain a sufficiently thin grid while keeping the computation time reasonable.

More precisely, we can illustrate how the algorithm works for a sender city $i$ when starting at a given day $d$. We initialize the step $t$ by $t = 0$ and the position $p_t$ by the position of the sender city:

- At step $t$, we extract from the input data the wind direction and speed information at day $d + t$, and we build the daily vector grid as described previously. Given that $p_t$ is not exactly at an intersection of the grid, nearest interpolation is used to approximate the wind vector at position $p_t$. Let's note that wind vector $w_t = (u_t, v_t)$.

- We look for potential receiver cities within a disk of radius $rad_t$ from $p_t$. For each city $r$ found, we define raw downwind intensity score as:

$$\text{Wind}_{i \to r, d, t} = \exp\{-\alpha \cdot rad_t - \beta \cdot |\theta|_{i \to r, d, t} - \gamma \cdot dist_{i \to r, d, t}\}$$

where $\alpha, \beta, \gamma$ are positive parameters.

The <u>first</u> component is the search radius at step $t$ ($rad_t$), which captures general decreases of downwind intensity oversteps. We increase the search radius by $0.2$ which represents about 21km at Brazil latitudes at each step to capture both the uncertainty in the streamline computation and the dispersion of air. The initial search radius $rad_0$ has a value of $2.8$ which represents about 300km at Brazil latitudes.

The <u>second</u> component enables to assign higher intensity to receiver cities that sit closer to the exactly-downwind direction of the sender. More precisely, let's note $l_{i \to r}$ the vector between sender city $i$ and receiver city $r$. The sense of the vector is not important. Let's note $\overline{v}_t$ the vector defined as $\overline{v}_t = (v_t, -u_t)/||w_t||$. Thus, it is a normal vector to the wind vector $w_t$ that has a norm of 1. Then, we define the absolute scalar product $|\theta|_{i \to r,d,t} = v_t \cdot l_{i \to r}$. The higher this term, the closer $l_{i \to r}$ is to be perpendicular to $w_t$. Since the higher $|\theta|_{i \to r,d,t}$, the lower the score $\text{Wind}_{i \to r,d,t}$, and thus the aim of this term is to penalize cities that are less impacted by the wind streamlines because they are not in the exactly-downwind direction from the wind streamline at that step. It is also important to note that $v_t$ is normalized to avoid seeing lower scores when the speed of the wind is higher. However, $l_{i \to r}$ is not normalized to penalize cities that are further from the sender city.

The <u>third</u> component is simply the distance between the sender and the current position $p_t$ ($\text{dist}_{i \to r,d,t}$), which captures geographic decay. This component is inspired by <u>Phillips et al. (2021)</u> that also uses an exponential decay with the distance from the emitter to model dispersion. This term also aims to penalize cities further from $p_t$ but gives more flexibility to the formula by decorrelating the decay in terms of angle (second term) and the one in terms of distance.

We assume $\text{Wind}_{i \to r,d,t}$ to be zero if $d_{i \to r,d,t} > \text{rad}_t$ (i.e., if receiver city lies outside of the search radius at step t) or if $\theta_{i \to r,d,t} > 0.4$ radian (i.e., if the receiver city is not obviously in the downwind direction from the wind streamline at that step). We choose parameter values $\{\alpha, \beta, \gamma\} = \{0.8, 0.49, 0.23\}$. These coefficients are chosen empirically so that the function that attributes wind scores over 7 days is approximately continuous. For that purpose, we used visualisations consisting in heatmaps that simulate the wind scores values not only for cities of interest but for all points of the map for different days and different sender cities. Examples of those heatmaps showing the approximate continuity of the wind score function for the final value of the parameters can be seen on the figure below.

- If $t < 6$, coefficients need to be updated for step $t + 1$. We increase $\text{rad}_t$ as described previously by $0.2$ to obtain $\text{rad}_{t+1}$. We update $p_t = (x_t, y_t)$ by following the local direction and speed of the wind i.e. using $w_t$:

$$x_{t+1} = x_t + 24 * 3600 * u_t / \text{dist}_m((x_t, y_t), (x_t + 1, y_t))$$

$$y_{t+1} = y_t + 24 * 3600 * v_t / \text{dist}_m((x_t, y_t), (x_t + 1, y_t))$$

$$p_{t+1} = (x_{t+1}, y_{t+1})$$

To understand those expressions, we must consider that $x$ and $y$ coordinates are in degrees while the vectors' coordinates $u$ and $v$ are in m/s. The distance (positive or negative) in meters crossed by the wind in 24 hours is of $d_{m,x} = 24 * 3600 * u_t$ along the x-axis and $d_{m,y} = 24 * 3600 * v_t$ along the y-axis. To obtain an approximation of the distance $d_p$ crossed in polar coordinates corresponding to a distance $d_m$ in meters, we use a cross product : if a delta of 1 degree in longitude at the latitude $y_t$ represents $\text{dist}_m((x_t, y_t), (x_t + 1, y_t))$ meters, then, an approximation of $d_p$ is $d_p \approx 1 * d_m / \text{dist}_m((x_t, y_t), (x_t + 1, y_t))$.

Thus,

$$d_{p,x} \approx 1 * d_{m,x} / \text{dist}_m((x_t, y_t), (x_t + 1, y_t))$$

$$d_{p,y} \approx 1 * d_{m,y} / \text{dist}_m((x_t, y_t), (x_t + 1, y_t))$$

hence the expression of $x_{t+1} = x_t + d_{p,x}$ and $y_{t+1} = y_t + d_{p,y}$.

We can then proceed to step $t + 1$.

Starting from each particular sender city and day of the period 1998-2001, we iterate the procedure for seven steps (i.e., a week) so for $t = 0$ to $t = 6$.

**Examples of wind indexes heat maps for two emitters**

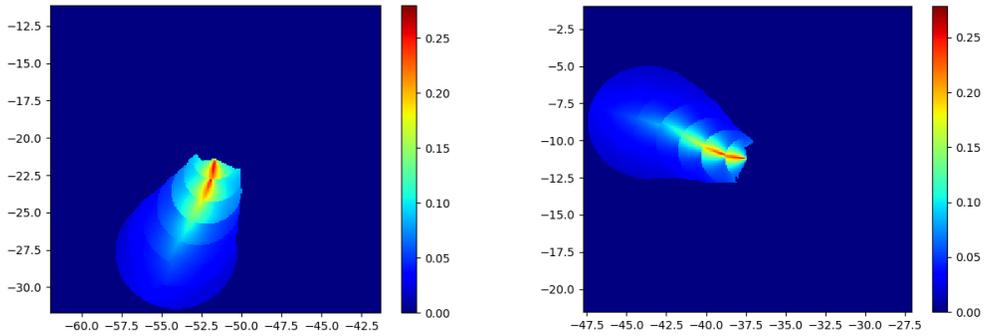

After the computation, we have a set of wind scores $\text{Wind}_{i \to r,d,t}$ that need to be aggregated at a day level, which means that we want to have a single coefficient for a given tuple (sender city $i$, receiver city $r$, date of arrival $d'$). For a given index a given index $\text{Wind}_{i \to r,d,t}$, the date of arrival $d'$ is the sum of the delay in days from the emission at the sender city i.e. step $t$ and of the date of emission $d: d' = t + d$. The downwind intensity score aggregated at a day-level for tuple (sender city $i$, receiver city $r$, date of arrival $d'$) is:

$$\text{Wind}_{i \to r,d'} = \sum_{d+t=d'} \text{Wind}_{i \to r,d,t}$$

After that, the second aggregation step is at a month-level to make the size of the regression dataset manageable in the econometric analysis. This aggregation consists in computing average intensity scores for each tuple (sender city $i$, receiver city $r$, month of arrival $m$):

$$\text{Wind}_{i \to r,m} = \text{Average}_{d' \in m} \text{Wind}_{i \to r,d'}$$

At the end, we record every pair (sender city $i$, receiver city $r$) among the monthly aggregated intensity scores. For each pair (sender city $i$, receiver city $r$), when there is no intensity score found for a given month $m$ of the period of interest (1998-2021), we add the monthly aggregated intensity score $\text{Wind}_{i \to r,m} = 0$. The final matrix containing monthly intensity scores should therefore present n rows per couple (sender city i, receiver city r) where n = 24 is the number of months in the period (1998-2021). Note that not every pair of cities would be in the matrix. Indeed, if wind "originating" from a city i has not reached city r within 7 days, for each day of the period as starting date, then no scores will be associated to couple (sender city i, receiver city r) in the summary matrix.